\documentclass[fleqn,usenatbib]{mnras}

\usepackage[T1]{fontenc}
\usepackage{ae,aecompl}
 
\usepackage{graphicx}	% Including figure files
\usepackage{amsmath}	% Advanced maths commands

\usepackage{amssymb}	% Extra maths symbols
\usepackage{float}
\usepackage{caption}
\usepackage{subcaption}

\title[Velocity structure in the Virgo cluster]{Measuring sloshing, merging and feedback velocities in the Virgo cluster} 

\author[Gatuzz et al.]{
Efrain Gatuzz$^{1}$\thanks{E-mail: egatuzz@mpe.mpg.de},
J. S. Sanders$^{1}$,
K. Dennerl$^{1}$,
C. Pinto$^{2}$,
A. C. Fabian$^{3}$,\newauthor
T. Tamura$^{4}$,
S. A. Walker$^{5}$
and J. ZuHone$^{6}$
\\
$^{1}$ Max-Planck-Institut f\"ur extraterrestrische Physik, Gie{\ss}enbachstra{\ss}e 1, 85748 Garching, Germany\\
$^{2}$ INAF - IASF Palermo, Via U. La Malfa 153, I-90146 Palermo, Italy \\
$^{3}$ Institute of Astronomy, Madingley Road, Cambridge CB3 0HA, UK\\ 
$^{4}$ Institute of Space and Astronautical Science (ISAS), Japan Aerospace Exploration Agency (JAXA) Kanagawa 252-5210, Japan\\ 
$^{5}$ Department of Physics and Astronomy, University of Alabama in Huntsville, Huntsville, AL 35899, USA\\
$^{6}$ Harvard-Smithsonian Center for Astrophysics, 60 Garden Street, Cambridge, MA, 02138, USA
}

\date{Accepted XXX. Received YYY; in original form ZZZ} 
\pubyear{2018} 
\begin{document}
 \label{firstpage}
\pagerange{\pageref{firstpage}--\pageref{lastpage}}
\maketitle 
\begin{abstract}
We present a detailed analysis of the velocity structure of the Virgo cluster using {\it XMM-Newton} observations. Using a novel technique which uses uses the Cu K$\alpha$ instrumental line to calibrate the EPIC-pn energy scale,  we are able to obtain velocity measurements with uncertainties down to $\Delta v \sim 100$ km/s. We created 2D projected maps for the velocity, temperature, metallicity, density, pressure and entropy with an spatial resolution of 0.25$'$. We have found that in the innermost gas there is a high velocity structure, most likely indicating the presence of an outflow from the AGN while our analysis of the cluster cool core using RGS data indicates that the velocity of the gas agrees with the M87 optical redshift. An overall gradient in the velocity is seen, with larger values as we move away from the cluster core. The hot gas located within the western radio flow is redshifted, moving with a velocity $\sim 331$ km/s while the hot gas located within the eastern radio flow is blueshifted, with a velocity $\sim 258$ km/s, suggesting the presence of backflows. Our results reveal the effects of both AGN outflows and gas sloshing, in the complex velocity field of the Virgo cluster.

\end{abstract} 

%%%% KEYWORDS %%%%


\section{Introduction}\label{sec_in}   
Measuring velocities of the intracluster medium (ICM) is important to understand the physical properties of such environment. Turbulent motions provide additional pressure support, particularly at large radii, which might affect calculations of hydrostatic equilibrium and cluster mass estimates \citep[e.g.][]{lau09}. Measurements of velocities can help to constrain active galactic nucleus (AGN) feedback models, given that how energy is distributed from AGN feedback within the bulk of the cluster depends on the balance between sound waves or shocks and turbulence \citep[see][for a review]{fab12b}. Moreover, the microphysics of the ICM, such as viscosity, can be probed by measuring velocities. Simulations predict that there is a close connection between entropy, temperature, density, pressure and velocity power spectra \citep[see e.g.][]{gas14,zhu14b,moh19}. Motions also transport of metals within the ICM, due to sloshing and uplift of metals by AGN \citep[e.g.][]{sim08}. In addition, velocities can directly indicate the sloshing of gas in cold fronts, which can remain for several Gyr \citep{asc06,roe12}. Despite its importance, the velocity structure of the ICM remains poorly constrained observationally. 

Measurements of line broadening and resonant scattering yield low turbulence motion, with velocities between $100-300$ km/s \citep{san10,san13,pin15,ogo17,liu19}. The {\it Hitomi} observatory \citep{tak16} directly measured random and bulk motions in the ICM using the Fe-K emission lines due to to its high spectral resolution microcalorimeter SXS X-ray detector. In the core of the Perseus cluster it measured a line-of-sight velocity dispersion of $164 \pm 10$ km/s between radii of $30$ and $60$ kpc and a gradient of $150$ km/s bulk flow across $60$ kpc of the cluster core \citep{hit16}. Despite the obvious impact of the AGN and sloshing of the surrounding ICM, these results showed that the Perseus core is not strongly turbulent. Such low level of turbulence is not fast enough to replenish heating in the short time necessary to balance cooling, although sound waves or shocks can plausibly do this \citep{fab17}. Simulation studies of Perseus-like clusters with either AGN \citep{lau17} or sloshing \citep{zuh18} show that the same level of low velocities can be achieved, even without including viscosity effects. Unfortunately, we will not be able to make further measurements with {\it Hitomi} in other clusters or in different regions of Perseus due to its loss. The next planned observatories with a high-resolution instrument capable of measuring velocities are {\sc XRISM} \citep{xri20}, to be launched in 2023 and {\sc Athena} \citep{bar17} from 2031.

\citet{san20} present a novel technique to perform velocity measurements using background X-ray lines seen in the spectra of the {\it XMM-Newton} EPIC-pn detector to calibrate the absolute energy scale of the detector to better than $150$ km/s at Fe-K. Using this technique, \citep{san20} mapped the bulk velocity distribution of the ICM over a large fraction of the central region of two nearby clusters of galaxies, Perseus and Coma. For the Perseus cluster, they detect evidence for sloshing associated with a cold front. In the case of the Coma cluster, they found that the velocity of the gas is close to the optical velocities of two central galaxies, NGC 4874 and NGC 4889, respectively.  
  
The cool-core Virgo/M87 cluster constitutes an excellent laboratory to apply such technique in order to study the ICM velocity structure. The AGN in its core is active, and displays large extended radio bubbles as well as a central jet \citep{owe00}, creating through feedback multiple shocks and large cavities in the ICM \citep{for07}.  The {\it XMM-Newton} observations of this system were analyzed by \citet{sim07}, showing that the X-ray emitting gas forms arms, tracing in the direction of these cavities, which coincide with powerful radio lobes. \citet{sim08} found that the abundance ratios of element such as O, Si, S and Fe are similar in and outside the X-ray arms, indicating that the metals must have been transported after the last major epoch of star formation. They estimated the mass of the gas with $kT_{e}<1.5$ keV to be about $5\times 10^{8}$ M$\odot$. \citet{sim10} identified two cold fronts, one towards the NW with a radius of $\sim 90$ kpc and one towards the SE at a radius of $\sim 33$ kpc, with sloshing of gas as the most viable explanation for the presence of both cold fronts. \citet{urb11} suggested that the unusually shallow density profile of the cluster, as well as the sharp drops in metallicity and temperature, are due to gas clumping, which increases as a function of radius. 

By applying a modified $\Delta$-variance data analysis method to {\it Chandra} observations of the cluster, \citet{zhu14} studied the ICM heating rate assuming that the ICM surface brightness fluctuations were due to turbulence. They concluded that turbulent heating is sufficient to offset radiative cooling, and even to balance it locally at each radius. \citet{wer16}, on the other hand, showed that the magnetic fields, aligned with the cold front, suppress the ICM diffusion conduction and mixing. Linear features were also observed, separated by $15$~kpc and around 10$~\%$ brighter than the surrounding gas, which may due to amplification of the magnetic fields by gas sloshing below the cold front. All these energetic phenomena are expected to produce motions which we aim to detect. 
 
In this paper, we present an analysis of the ICM velocity structure within the Virgo galaxy cluster. The outline of this paper is as follows. In Section~\ref{sec_dat} we describe the data reduction process.  In Section \ref{sec_fits} we explain the fitting procedure. A discussion of the results is shown in Section~\ref{sec_dis} while the conclusions and summary are included in Section~\ref{sec_con}. Throughout this paper we assumed the distance of Virgo to be the cosmological distance given by the redshift of the central galaxy M87 \citep[$z=0.00436\pm 0.000023$][]{all14} and a concordance $\Lambda$CDM cosmology with $\Omega_m = 0.3$, $\Omega_\Lambda = 0.7$, and $H_{0} = 70 \textrm{ km s}^{-1}\ \textrm{Mpc}^{-1} $.

\section{Data reduction}\label{sec_dat}
\subsection{XMM-newton}

\begin{table} 
\footnotesize
\caption{\label{tab_obsids}{\it XMM-Newton} observations of the Virgo cluster.}
\centering 
\begin{tabular}{cccccccc}   
\hline
ObsID & RA & DEC & Date & Exposure \\
 &&&Start-time& (ks)\\
\hline
0106060101	&	187.70	&	12.72	&	2001-07-12  &	9.8	    \\
0114120101	&	187.71	&	12.36	&	2000-06-19 	&	60.1	\\
0200920101	&	188.07	&	12.26	&	2005-01-10 	&	109.3	\\
0551870101	&	188.07	&	12.55   &	2008-06-18  &	19.9	\\
0551870301	&	187.48	&	12.11	&	2008-06-20  &	26.5	\\
0551870401	&	187.83	&	12.06	&	2008-06-17  &	21.6	\\
0551870501	&	187.36	&	12.43	&	2008-06-26  &	38.9	\\
0551870601	&	187.52	&	12.73	&	2008-06-19  &	21.3	\\
0551870701	&	188.07	&	12.26	&	2008-06-30  &	26.6	\\
0551871201	&	187.83	&	12.76	&	2008-12-06  &	8.9  	\\
0551871301	&	187.46	&	12.11	&	2008-12-11  &	30.6	\\
0603260201	&	187.50	&	12.71	&	2009-07-08  &	17.9	\\
0603260401	&	187.81	&	12.76	&	2009-07-07  &	12.9	\\
0603260501	&	188.07	&	12.24	&	2009-06-04  &	12.9	\\
0603260601	&	187.70	&	12.39	&	2009-06-19  &	20.0	\\
0803670501	&	187.70	&	12.39	&	2017-07-06  &	132.0	\\
0803670601	&	187.70	&	12.39	&	2017-06-16  &	65.0	\\
0803671001	&	187.63	&	12.44	&	2017-12-15  &	63.0	\\
0803671101	&	187.63	&	12.44	&	2017-12-25  &	131.9	\\
0823270101	&	187.39	&	12.63	&	2018-06-09  &	24.0    \\
\end{tabular}
% \vspace{-7mm}
\end{table}

Table~\ref{tab_obsids} shows the specifications, including IDs, dates and clean exposure times, of the Virgo cluster observations analyzed in this paper. 

  \begin{figure*}
        \centering
        \begin{subfigure}{0.55\textwidth}
            \includegraphics[scale=0.76]{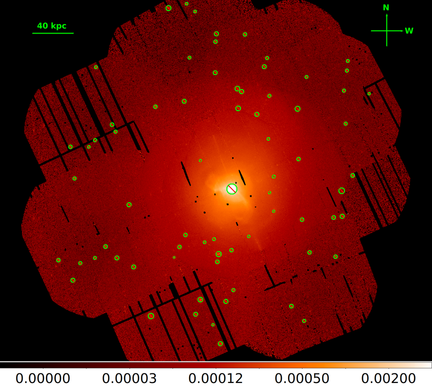}
        \end{subfigure}
          \centering
        \hspace{1.8cm}          
        \begin{subfigure}{0.32\textwidth}
        	\vspace{-0.45cm}
           \includegraphics[scale=0.368]{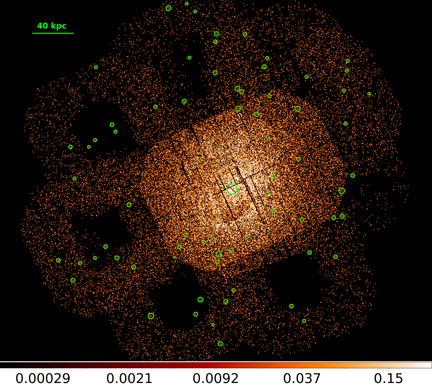}\\
           \includegraphics[scale=0.368]{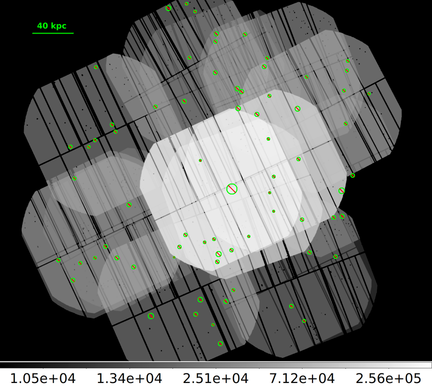}
        \end{subfigure}
        \caption{\emph{Left panel:} X-ray surface brightness of the Virgo cluster in the 0.5 to 9.25 keV energy range. Green circles indicate the point-like sources, as well as the AGN in the cluster core, which were excluded in the analysis. \emph{Top right panel:} Fe-K count map, showing the number of counts in each 1.59 arcsec pixel in the 6.40-6.90 keV Fe-K complex and excluding the Cu holes for which the new EPIC-pn energy calibration scale cannot be applied. \emph{Bottom right panel:} total exposure time (s) in the 4.0-9.25 keV energy range. } \label{fig_xray_maps} 
    \end{figure*}

\subsubsection{EPIC-pn energy scale calibration}\label{sec_dat_cal}
The {\it XMM-Newton} European Photon Imaging Camera \citep[EPIC,][]{str01} spectra were reduced with the Science Analysis System (SAS\footnote{\url{https://www.cosmos.esa.int/web/xmm-newton/sas}}, version 18.0.0). First, we processed each observation with the {\tt epchain} SAS tool. We used only single-pixel events (PATTERN==0) while bad time intervals were filtered from flares applying a 1.0 cts/s rate threshold. In order to avoid regions close to CCD edges or bad pixels we filtered the data using FLAG==0. 

\citet{san20} calibrated the absolute energy scale of the EPIC-pn detector to better than $150$ km/s at Fe-K by using background X-ray lines seen in the spectra of the detector. In order to account for and correct gain variation they measured the Fe-K line centroid relative to the strong Cu K$\alpha$ instrumental line. Because these lines are close in energy, large energy extrapolations are avoided. The calibration process includes corrections for the average gain of the detector during the observation (first step), for the spatial gain variation across the detector with time (second step) and the final correction to the energy scale as function of detector position and time (third step). The observations analyzed in \citet{san20} to perform the calibration were chosen to cover the full mission lifetime from 2000 to 2018. It is important to note that, in order to study velocity structures in cluster cores, observations must be taken offset because of the ring-like structure of the Cu K$\alpha$ line, caused by the electronics board.

We have performed a new calibration of the EPIC-pn absolute energy scale by including observations taken in the last two year, corresponding to 33 in Full Frame mode and 42 in Extended Full Frame mode new observations \citep[see][ Section 2 for a complete description of the calibration process]{san20}. We noted that the SAS version 18.0.0 includes corrections for the EPIC-pn energy scale, compared to the version 16.1.0 used in \citet{san20}. In this sense, we have redone the analysis of the Perseus cluster shown in \citet{san20} with the new calibration files and we have found that there are no significant changes in the results. Then, we applied the energy calibration scale correction to the event files obtained for the Virgo cluster observations listed in Table~\ref{tab_obsids}. 

Figure~\ref{fig_xray_maps} left panel shows the X-ray image obtained in the 0.5-2.0 keV energy band. The image shows the gas arm structures, which trace the direction of the feedback large cavities produced by the AGN central jet. Identification of point sources was performed using the SAS task {\tt edetect\_chain}, with a likelihood parameter {\tt det\_ml} $> 10$. The point sources were excluded from the subsequent analysis, including the AGN in the cluster core (i.e. a central circular region with a diameter  $D=58$ \arcsec). Figure~\ref{fig_xray_maps} top right panel shows the number of counts in each 1.59 arcsec pixel in the Fe-K complex (6.50 to
6.90 keV, restframe), after subtracting neighbouring scaled continuum images (6.06 to 6.43 and 6.94 to 7.16 keV, restframe). Gaussian smoothing of $\sigma = 4$ pixels was applied. Due to the spatial offset in the observations we have a large number of counts in multiple directions around the cluster center. Figure~\ref{fig_xray_maps} bottom right panel shows the total exposure time in the 4.0-9.25 keV energy range.

\subsubsection{RGS}\label{sec_dat_rgs}
High resolution spectra from the Reflection Grating Spectrometers (RGS) on {\it XMM-Newton} data provide an alternative way to measure the redshift of the X-ray emission close to the nucleus. The RGS instruments have a softer energy range than the pn detector ($\sim0.5$ to $2$ keV) and so are sensitive to cool X-ray emitting components. With the RGS instruments the obtained spectral resolution reduces as the extent of the source on the sky increases. The extraction region is a one-dimensional slit across the object, the orientation of which depends on the roll angle of the telescope. These constraints mean that the RGS data are only sensitive to cool material in the cluster core and are limited to single measurements.

For each of the four observations with the nucleus in the field of view we extracted the spectrum from $90$ per cent of the point spread function (PSF) width including $95$ per cent of the pulse height distribution, assuming the nucleus lies at a position of Very-long-baseline interferometry (VLBI) measurements \citep[RA=187.705, Dec=12.391, ][]{fey04}. We created response matrices with 32000 bins in wavelength (to properly sample the velocity space), combining the spectra from the two RGS detectors.

\subsection{Chandra}\label{sec_dat_cha}
\begin{figure}    
\centering
\includegraphics[width=0.47\textwidth]{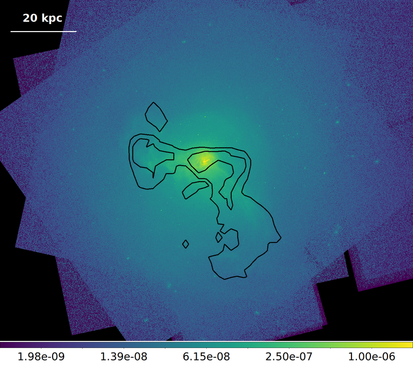}
\caption{{\it Chandra} image of the Virgo cluster in the 0.5--7.0 keV energy range. The 3-levels contour map from the radio data are included as reference for the scale.} \label{fig_chandra1} 
\end{figure}

In order to compute thermodynamic profiles, we reprocessed the {\it Chandra} ACIS-I observations of the core of M87 (OBSIDs 5826, 5827, 5828, 6186, 7210, 7211 and 7212) using {\tt acis\_process\_events}, applying {\tt vfaint} filtering where appropriate. Figure~\ref{fig_chandra1} shows the {\it Chandra} image in the 0.5--7 keV energy range with  the 3-levels contour maps from the radio observations from \citet{owe00} superposed as reference for the size scale.  Lightcurves in the 0.5 to 10 keV band were made from each observation, using CCDs 0, 1 and 2 in 200s time bins, excluding point sources detected using the CIAO {\tt wavdetect} tool. Sigma clipping was iteratively applied, removing periods outside $2.5\sigma$ from the median, to produce a set of good time intervals which were then used to filter the data. Standard blank-sky background datasets were obtained for each observation. The exposure times of the datasets were adjusted so that the background rate in the 9-12 keV band matched that of the respective source datasets. The background dataset lengths were adjusted (by discarding random events and modifying exposure), so that the ratio of a background exposure to the total background exposure matched the respective ratio of the source exposure to the total source exposure. The backgrounds were reprojected to match each respective observation. The observations and background datasets were finally reprojected to a common aimpoint.

To detect point sources we applied the {\tt wavdetect} tool to find sources using wavelet scales of 2 and 4 to images in the 0.5 to 7.0 keV band using binning of 1 sky pixel (0.5 arcsec). These sources were manually edited, to remove sources associated with the extended X-ray emission in the core of the cluster and to include any missed due to this emission. An image mask for these sources was created for our analysis. We created images, background-images and exposure maps of the cluster in the energy bands 0.50--0.80, 0.80--1.20, 1.20--1.55, 1.55--2.07, 2.07--3.00, 3.00--4.00, 4.00--5.00, 5.00--6.00 and 6.00--7.00 keV (assuming monochromatic exposure maps at the midpoints of these bands). As the nucleus is bright, the columns for each observation associated with the nucleus were masked-out to reduce the effect of out-of-time events. Using these masks we computed total images and total exposure maps. We created an average response matrix and ancillary response matrix for the observations using a circular region located at the M87 nucleus with a 9.8 arcsec radius.

\begin{figure}   
\centering 
\includegraphics[width=0.35\textwidth]{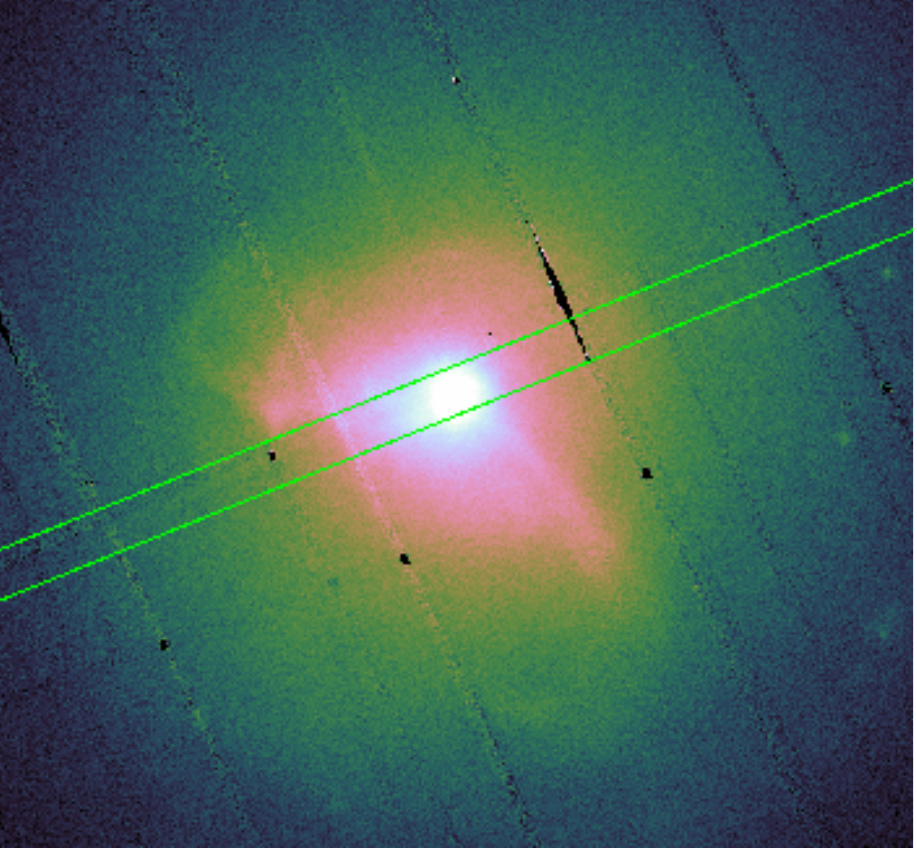}\\
\includegraphics[width=0.47\textwidth]{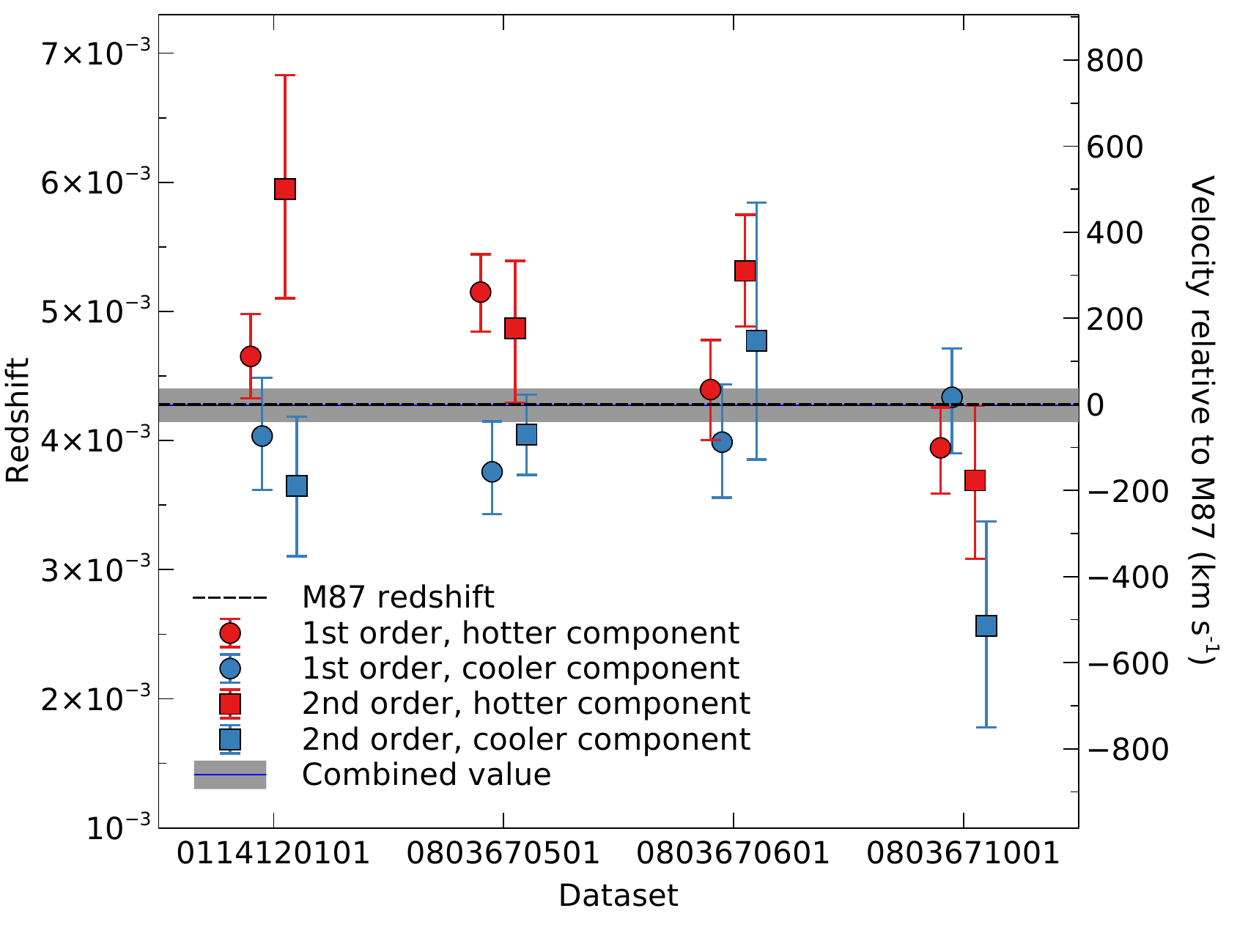}
\caption{\emph{Top panel:} RGS extraction region used for fitting the M87 redshift. \emph{Bottom panel:} Best fitting redshifts obtained from the RGS data. The independent sets of redshifts obtained from each of four of XMM RGS observations of M87 are plotted. For each observation the first and second order spectra were separately fit using two components with independent temperatures and redshifts. The M87 optical redshift is also shown with the 1$\sigma$ range of the redshift which best fits the data points.}\label{fig_rgs} 
\end{figure}

\begin{figure}   
\centering 
\includegraphics[width=0.48\textwidth]{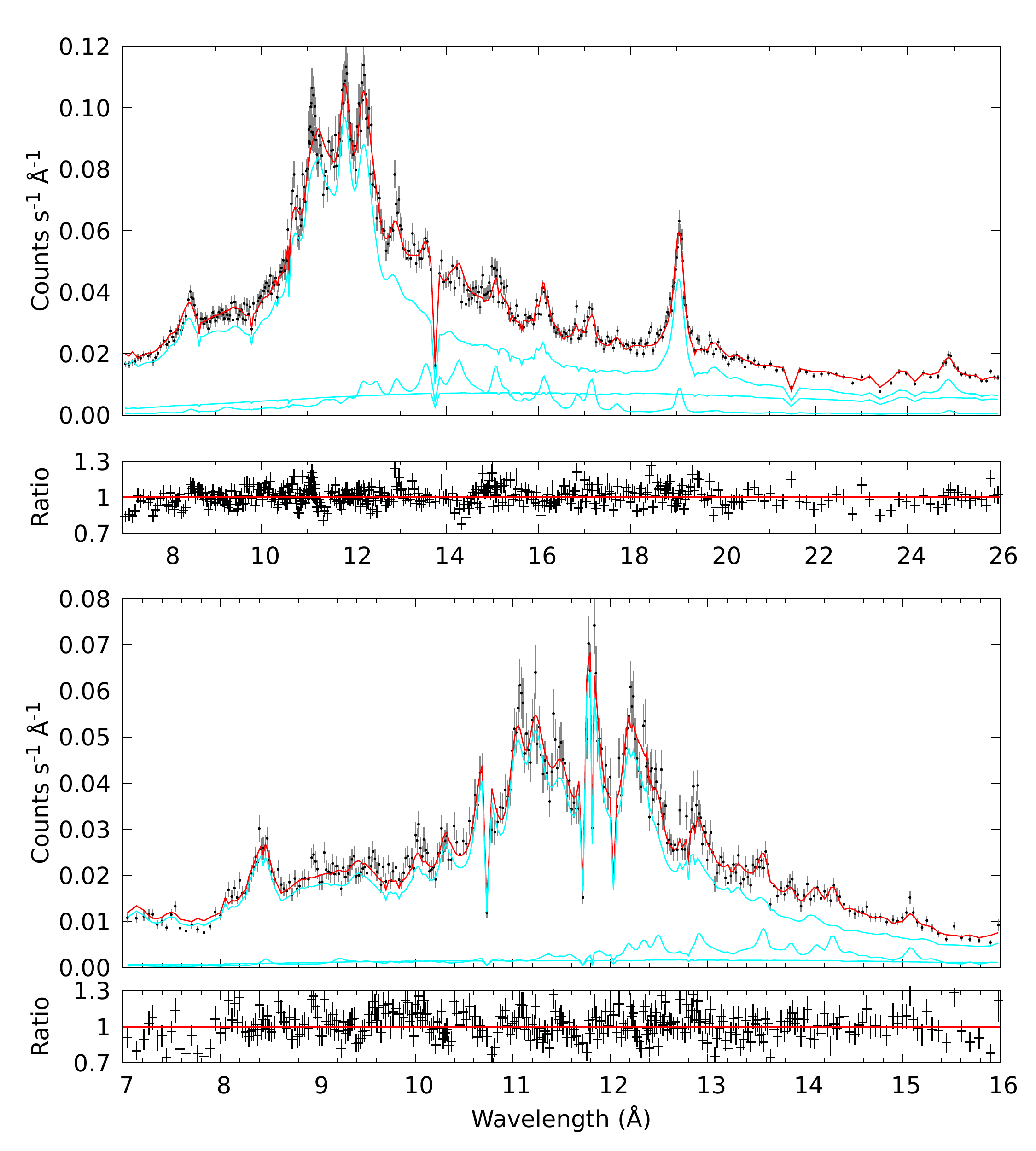}
\caption{RGS spectra for OBSID 0803671101. The 1st order spectrum (top panel) was rebinned to S/N=15 and the 2nd order spectrum (top panel) to S/N=10 for illustrative purposes.}\label{fig_rgs_spectra} 
\end{figure}

\section{Data analysis}\label{sec_fits} 
We model the cluster emission with the {\tt apec} spectral code \citep{fos19} version 3.0.9, which takes into account corrections derived from the {\it Hitomi} observations \citep{hit18}.  In order to account for the Galactic X-ray absorption we included a {\tt tbabs} component \citep{wil00}. The spectra were fitted in the 4.0-10 keV energy band. For each spatial region, we stacked the spectra from different observations together, using the number of counts in the fitting spectral region as a weighting factor. The free parameters in the model are the temperature, metallicity (i.e. metal abundances with He fixed at cosmic), redshift and normalization. We assumed a column density of $1.27\times 10^{20}$ cm$^{-2}$ \citep{kal05}, although in the energy range analyzed the absorbing component has a weak effect in the modeling. For the X-ray spectral fits we use the {\it xspec} data fitting package (version 12.10.1\footnote{\url{https://heasarc.gsfc.nasa.gov/xanadu/xspec/}}). We also perform model fits using the {\tt spex} spectral code \citep{kaa96} instead of {\tt apec} by importing its collisional ionization equilibrium model into {\it xspec} and using version 3.06 of its atomic tables and code. We assumed {\tt cash} statistics \citep{cas79}. Errors are quoted at the 1-$\sigma$ confidence level unless otherwise stated. Finally, abundances are given relative to \citet{lod09}. As background components we have included Cu-$K\alpha$, Ni-$K\alpha$, Zn-$K\alpha$ and Cu-$K\beta$ instrumental emission lines, and a powerlaw component with its photon index fixed at 0.136 \citep[the average value obtained from the archival observations analyzed in ][]{san20}.

\section{Results and discussion}\label{sec_dis} 

\subsection{RGS spectral analysis}\label{sec_rgs}

In order to measure the M87 cluster center velocity we fitted the 1st and 2nd order spectra separately between 7-26\AA, and 7-16\AA, respectively. Figure~\ref{fig_rgs}, top panel, shows the RGS extraction region. It is important to note that we excluded the cluster center from our EPIC-pn analysis and therefore a direct comparison with the RGS measurement cannot be done. Each spectrum was fitted by a two-component {\tt vapec} model with individual redshifts, temperatures and extent model, absorbed by a {\tt TBabs} model\footnote{Note that the RGS energy range allows us to include two {\tt vapec}  components to model the cool X-ray emitting plasma while for the EPIC-pn we only include one component.}. The metallicities were linked between the two components. For the first-order spectra, the He metallicity was fixed at solar and the C, Al, Si and S metallicities were tied to Fe. For the second-order spectra Ar and Ca were also tied to the Fe metallicity. The extent model applied a variable Gaussian broadening in wavelength space to account for source spatial extent \citep[see ][]{bri98}. We also included an absorbed powerlaw with a free photon index which is allowed to vary in the fits to model the central AGN. As an example, Figure~\ref{fig_rgs_spectra} shows the spectra and best-fitting model for the observation 0803671101.

Figure~\ref{fig_rgs}, bottom panel, shows the best fitting redshifts for each of the observations, each of the orders and each of the temperature components, compared to the optical velocity of M87. If we assume the data points are independent, drawn from a Gaussian distribution and have the correct statistical uncertainties, we obtain a mean redshift of $0.00427\pm 0.00013$ and width $\sigma=0.00037\pm0.00016$. The spread in velocity is consistent with the expected wavelength scale accuracy of the instrument. We found that the M87 velocity obtained from this analysis is in good agreement with the optical velocity of M87.

  \begin{figure}
   \centering
\includegraphics[width=0.46\textwidth]{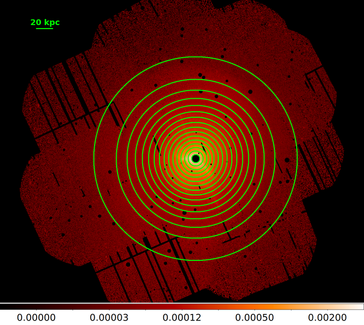}\\
\includegraphics[width=0.46\textwidth]{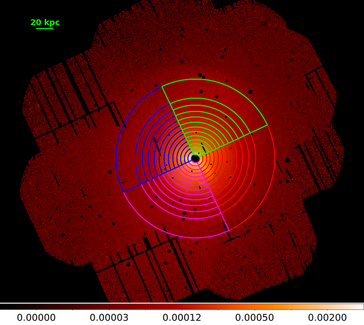}
\caption{\emph{Top panel:} Virgo cluster extracted regions for Case 1. \emph{Bottom panel:} Virgo cluster extracted regions for Case 2. Four zones are defined: N (green), W (red) S (magenta) and E (blue). In both figures black circles correspond to point sources which were excluded from the  analysis, including the AGN in the cluster core.} \label{fig_cas12_region} 
    \end{figure}

    \begin{figure*}
   \centering
\includegraphics[width=1.0\textwidth]{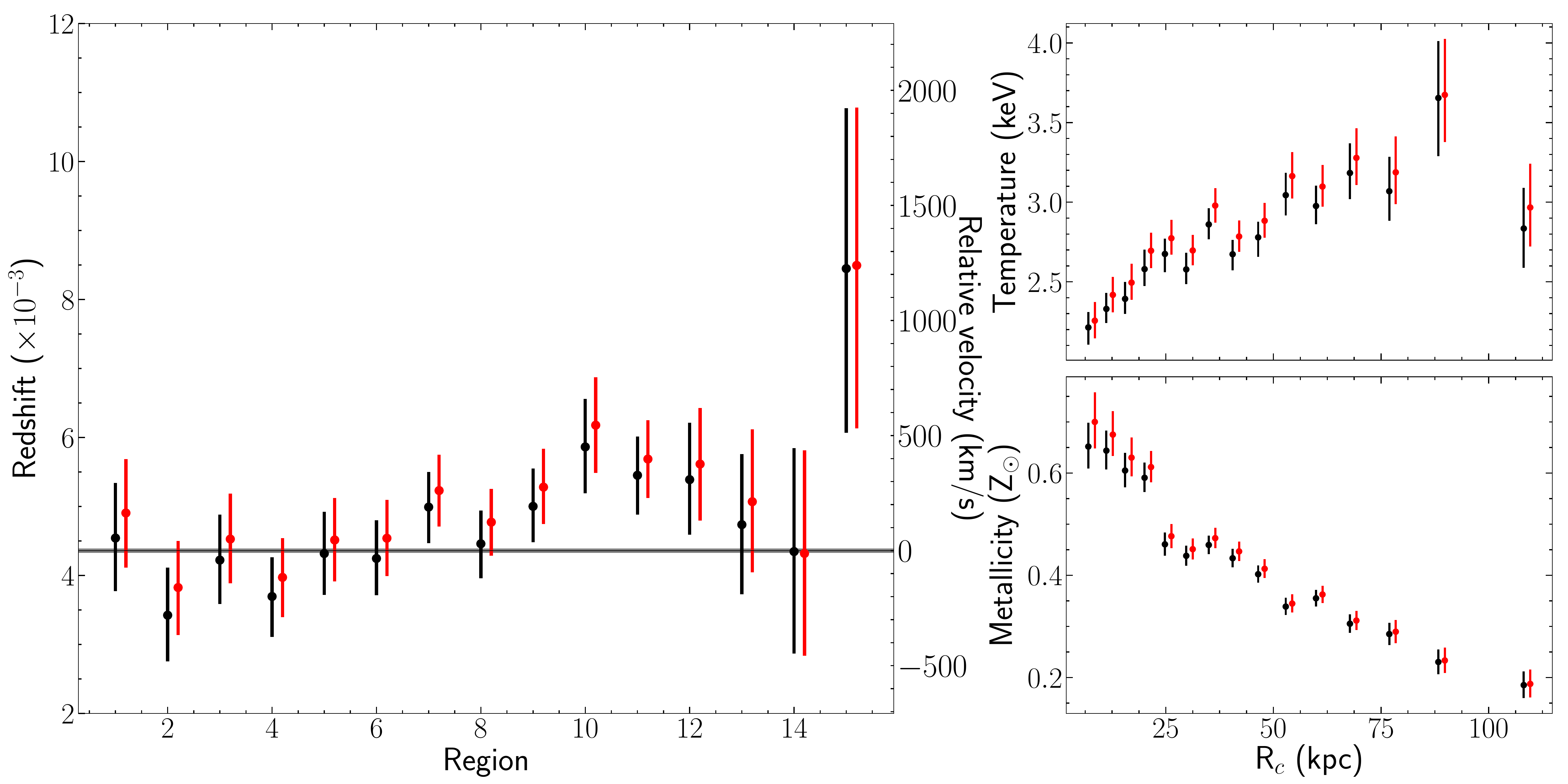}
        \caption{\emph{Left panel:} velocities obtained for each region for Case 1 (numbered from the center to the outside). The M87 redshift is indicated with an horizontal line. \emph{Right panels:} temperature and metallicity profiles obtained from the best fit results. Black points correspond to results obtained with the {\tt apec} model while red points correspond to {\tt spex} CIE model. } \label{fig_cas1_results} 
    \end{figure*}

\begin{table}
\scriptsize
\caption{\label{tab_cas1}Virgo cluster best-fit parameters: Case 1. }
\centering
\begin{tabular}{ccccccc}
\\
Region &\multicolumn{5}{c}{{\tt apec} model}  \\
\hline
 &$kT$& Z& $z$   & $norm$   & cstat/dof\\ 
  & &  &  ($\times 10^{-3}$) &   ($\times 10^{-3}$)  \\ 
\hline
\hline
\\  
1 &$2.21\pm 0.11$&$0.65\pm 0.05$&$4.54_{-0.77}^{+0.80}$&$15.70_{-0.98}^{+1.28}$&$1026/1040$\\
2 &$2.33\pm 0.10$&$0.64\pm 0.04$&$3.42\pm 0.68$&$19.90\pm 1.19$&$1035/1099$\\
3 &$2.39\pm 0.10$&$0.60\pm 0.03$&$4.22\pm 0.65$& $8.71\pm 0.53$&$1133/1143$\\
4 &$2.58\pm 0.12$&$0.59\pm 0.03$&$3.70\pm 0.58$& $9.29\pm 0.58$&$1118/1170$\\
5 &$2.67\pm 0.11$&$0.46\pm 0.02$&$4.32\pm 0.60$&$10.43_{-0.49}^{+0.68}$&$1266/1182$\\
6 &$2.58\pm 0.10$&$0.44\pm 0.02$&$4.25\pm 0.54$&$14.24\pm 0.79$&$1164/1188$\\
7 &$2.86\pm 0.10$&$0.46\pm 0.02$&$4.99\pm 0.52$&$15.54\pm 0.63$&$1225/1189$\\
8 &$2.67\pm 0.10$&$0.43\pm 0.02$&$4.46\pm 0.49$&$22.00_{-0.95}^{+1.25}$&$1223/1189$\\
9 &$2.78\pm 0.11$&$0.40\pm 0.02$&$5.00_{-0.52}^{+0.55}$&$23.90_{-0.96}^{+1.43}$&$1300/1189$\\
10&$3.04\pm 0.14$&$0.34\pm 0.02$&$5.86\pm 0.68$&$19.82\pm 1.03$&$1248/1189$\\
11&$2.98\pm 0.13$&$0.36\pm 0.02$&$5.45\pm 0.57$&$22.38\pm 1.07$&$1363/1189$\\
12&$3.18_{-0.16}^{+0.19}$&$0.31\pm 0.02$&$5.39\pm 0.81$&$24.58\pm 1.57$&$1341/1189$\\
13&$3.07_{-0.19}^{+0.22}$&$0.29\pm 0.02$&$4.74\pm 1.02$&$35.25\pm 2.67$&$1358/1189$\\
14&$3.65\pm 0.36$&$0.23\pm 0.02$&$4.35\pm 1.49$&$23.73_{-1.95}^{+2.63}$&$1259/1189$\\
15&$2.83\pm 0.26$&$0.19\pm 0.03$&$8.45_{-2.38}^{+2.32}$&$37.06_{-3.61}^{+4.77}$&$1270/1189$\\
\\ 
 \hline
\end{tabular}
\end{table}

\subsection{Fitting spectra from concentric regions}\label{circle_rings}
 We proceed to analyze non-overlapping circular regions in order to study in detail the velocity structure. The thickness of these rings increases as the square root with distance from the M87 center. We have two different region extraction cases for the analysis:
 
\begin{itemize}
\item \emph{Case 1:} complete concentric rings with the center located in the cluster center, the central galaxy M87. 
\item \emph{Case 2:} concentric rings but divided in four zones (N, S, E and W).
\end{itemize}
    
Figure~\ref{fig_cas12_region} top panel shows the chosen regions for \emph{Case 1}. These regions are numbered from the center to the outside. For each zone we perform spectral fits using extracted spectra from all observations combined. Table~\ref{tab_cas1} lists the best-fit results. Figure~\ref{fig_cas1_results} left panel shows the complex velocity structure obtained. We have obtained accurate velocity measurements down to $\Delta v\sim 154$ km/s (for ring 7). The largest redshift/blueshift (i.e. with respect to M87) correspond to $1226_{-714}^{+697}$ km/s and $-280_{-200}^{+206}$ km/s for rings 15 and 2, respectively. We note that the inner regions tend to display blueshift while outer regions tend to display redshift. In this sense the average velocities for regions 1-6 is $-86_{-190}^{+193}$ km/s while the average velocities for region 7-14 is $201_{-227}^{+230}$ km/s. Moreover, the blueshifted material has lower temperatures ($<2.7$ keV) except for the innermost ring, which displays a velocity consistent with the rest frame and  has the lowest temperature ($=2.1\pm 0.11$ keV). The redshifted material, on the other hand, has higher temperatures ($>2.7$ keV). 

\begin{figure}    
\includegraphics[width=0.48\textwidth]{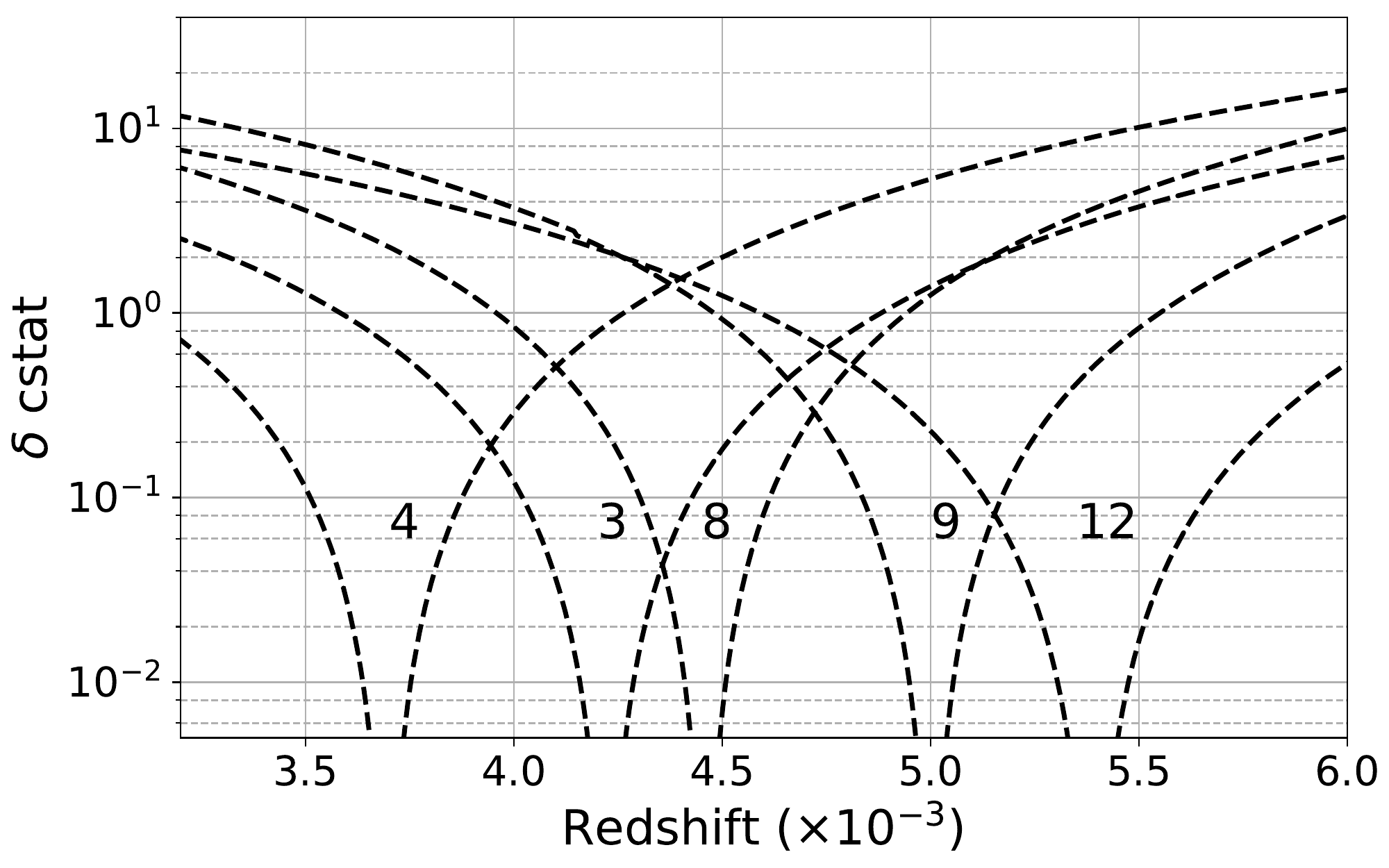}
\caption{Change in fit statistic as a function of redshift for selected regions in the Case 1 analysis, for the total spectral fit. The region numbers, with respect to Table~\ref{tab_cas1}, are indicated. } \label{fig_cas1_cstat} 
\end{figure}

Figure~\ref{fig_cas1_results} middle and right panels show the temperature and metallicity profiles obtained from the fits as a function of the radius. We have found a good agreement between the {\tt apec} (black points) and the {\tt spex} CIE (red points) best-fit results. Both profiles displays a discontinuity at $\sim 30$ kpc. \citet{sim07} discussed the presence of an inner cold front at such distance. Figure~\ref{fig_cas1_cstat} shows the change in the fit statistic if the redshift parameter changes from the best-fitting value and the spectra is refitted, for some selected spatial regions from Table~\ref{tab_cas1}. The curves are smooth and only display a single minimum, becoming asymmetric far from the best-fitting redshifts. This indicates that the redshift values obtained, as well as the statistical uncertainties, are robust. The overall gradient in the velocities, with larger values as we move away from the cluster core, may indicate the effects of gas sloshing.

\begin{figure*}  
\includegraphics[width=0.47\textwidth]{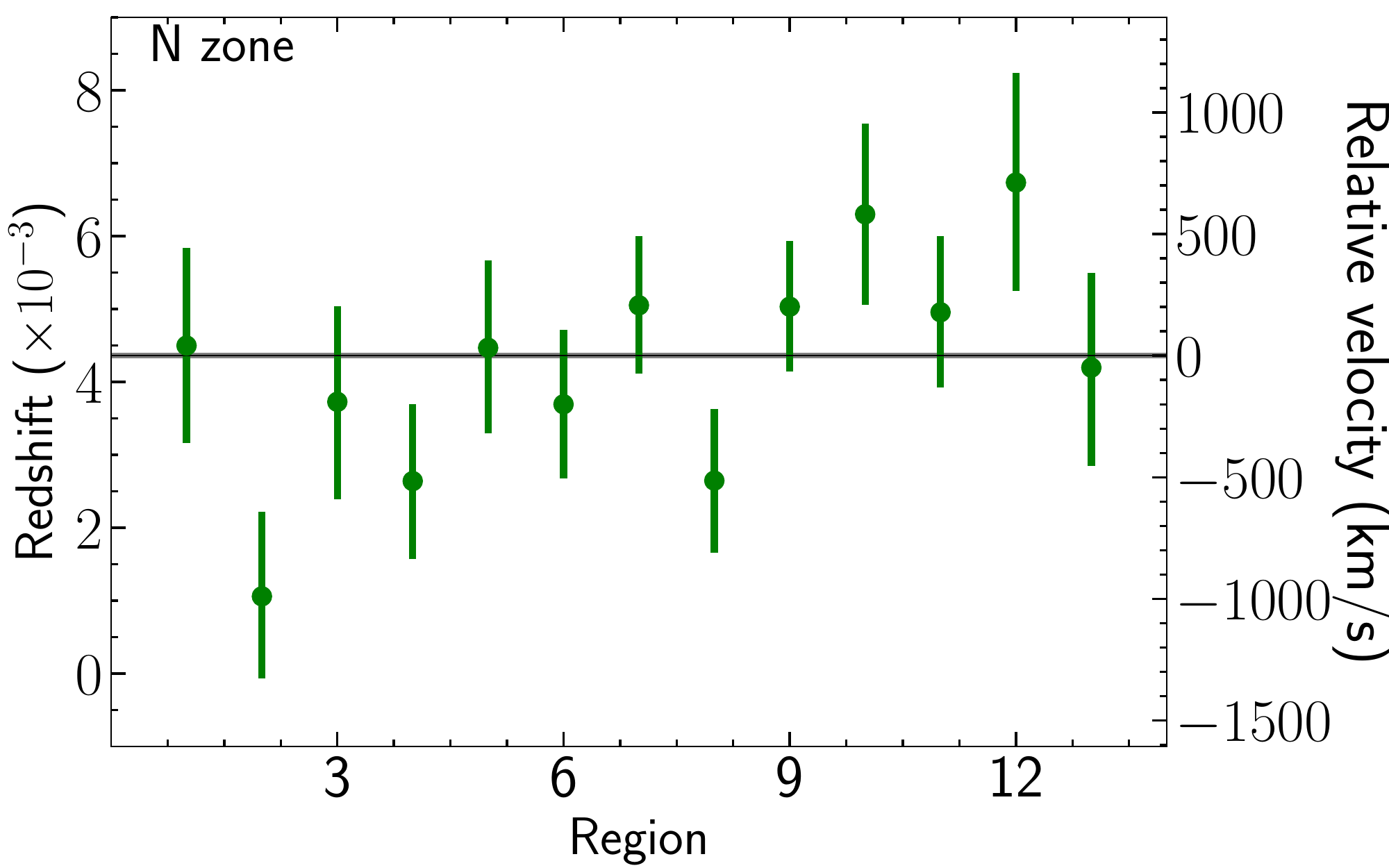}   
\includegraphics[width=0.47\textwidth]{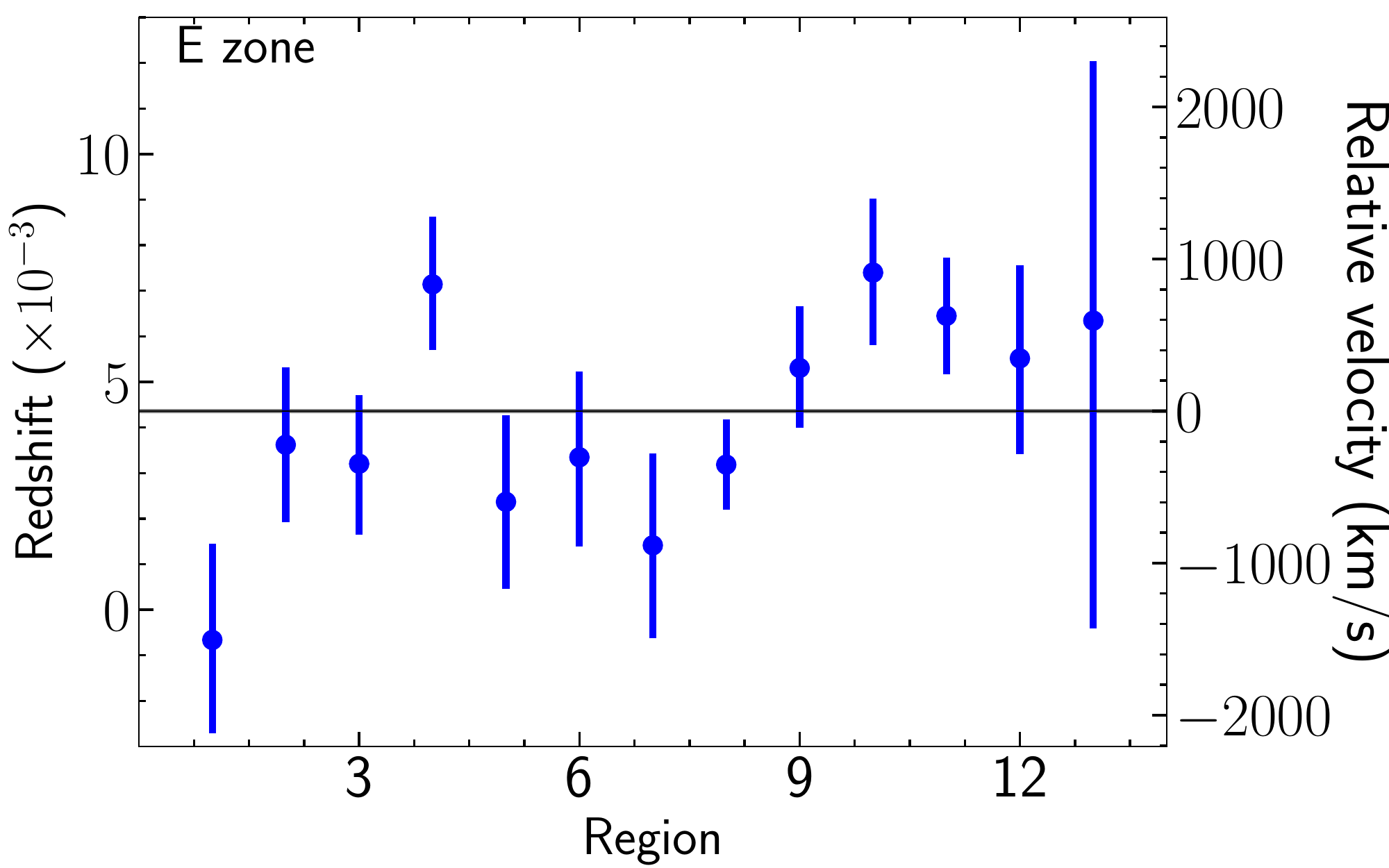}   \\
\includegraphics[width=0.47\textwidth]{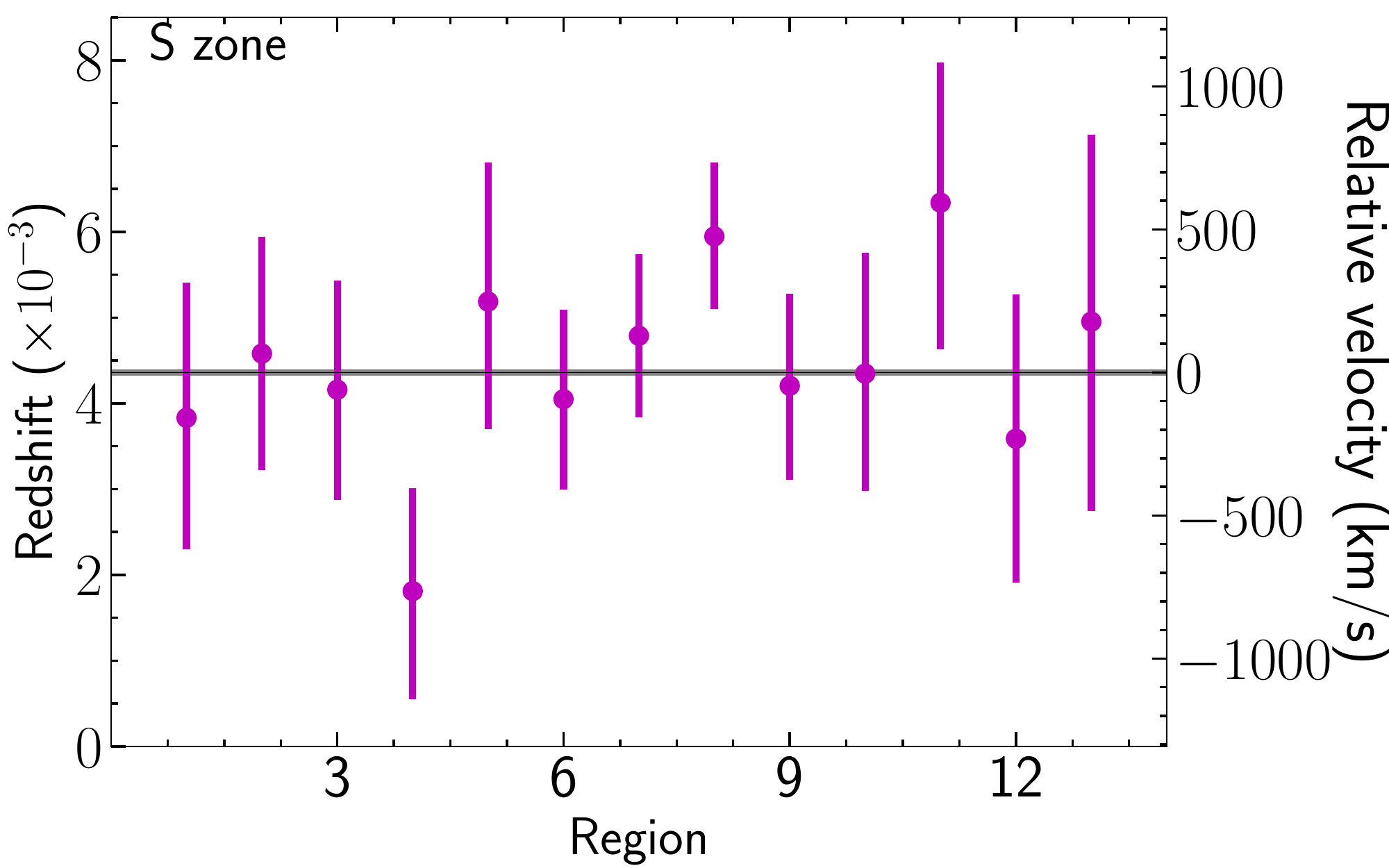}   
\includegraphics[width=0.47\textwidth]{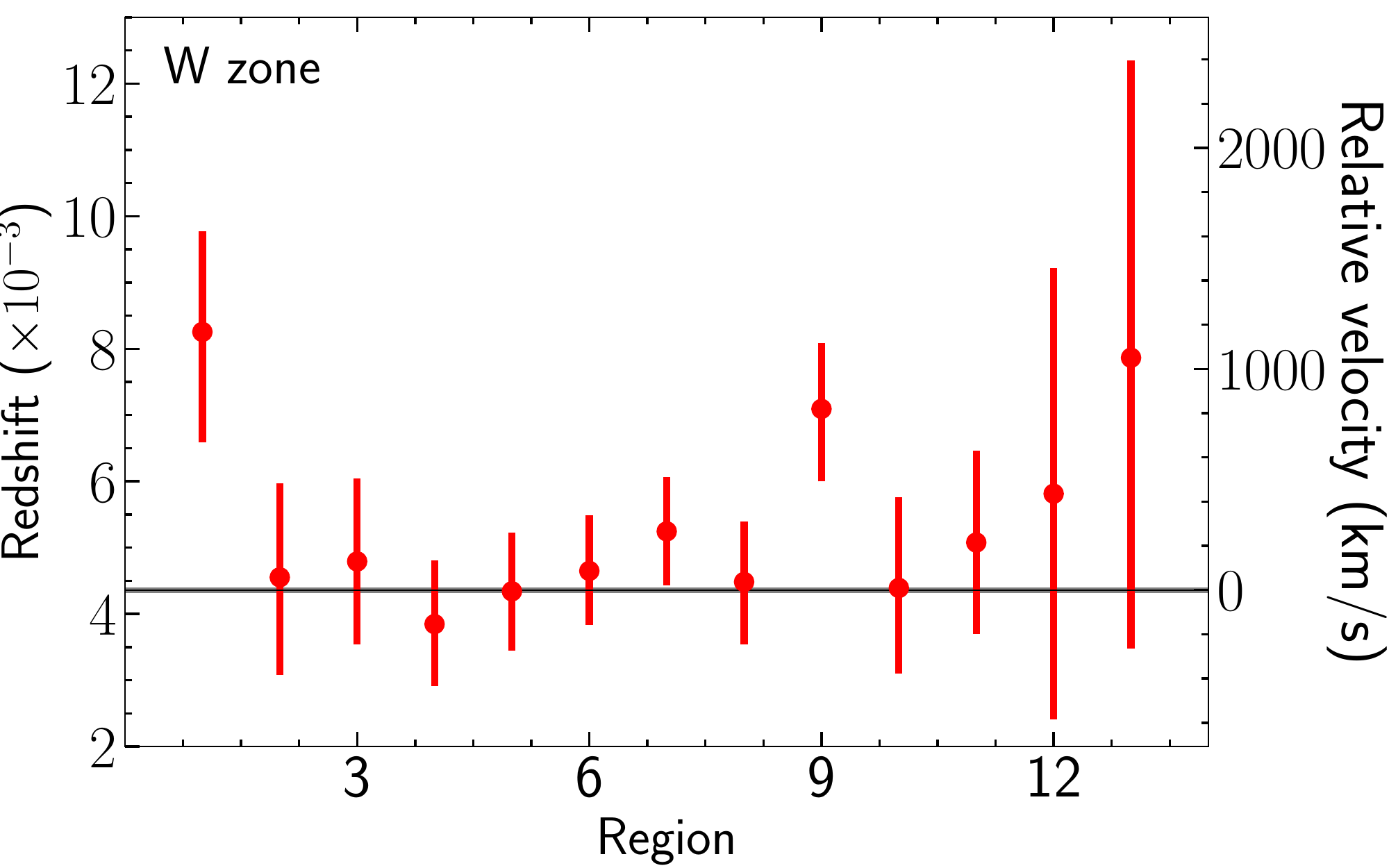}   
\caption{Velocity structure obtained for the Virgo cluster Case 2 analysis. The colors indicate the N (green), W (red) S (magenta) and E (blue). The M87 redshift is indicated with horizontal lines.} \label{fig_cas2_vel} 
\end{figure*}

\begin{figure*}  
\includegraphics[width=0.47\textwidth]{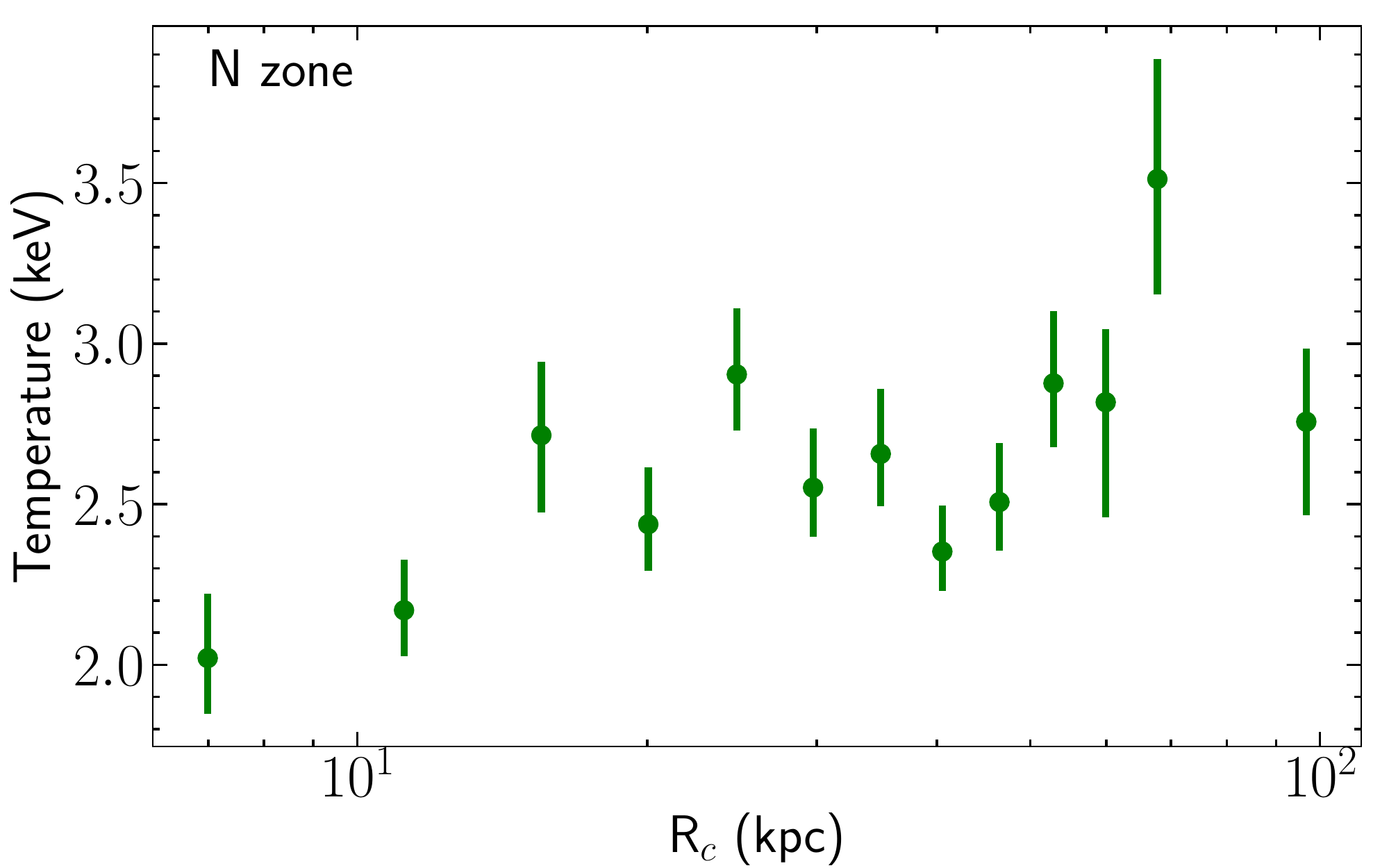}   
\includegraphics[width=0.47\textwidth]{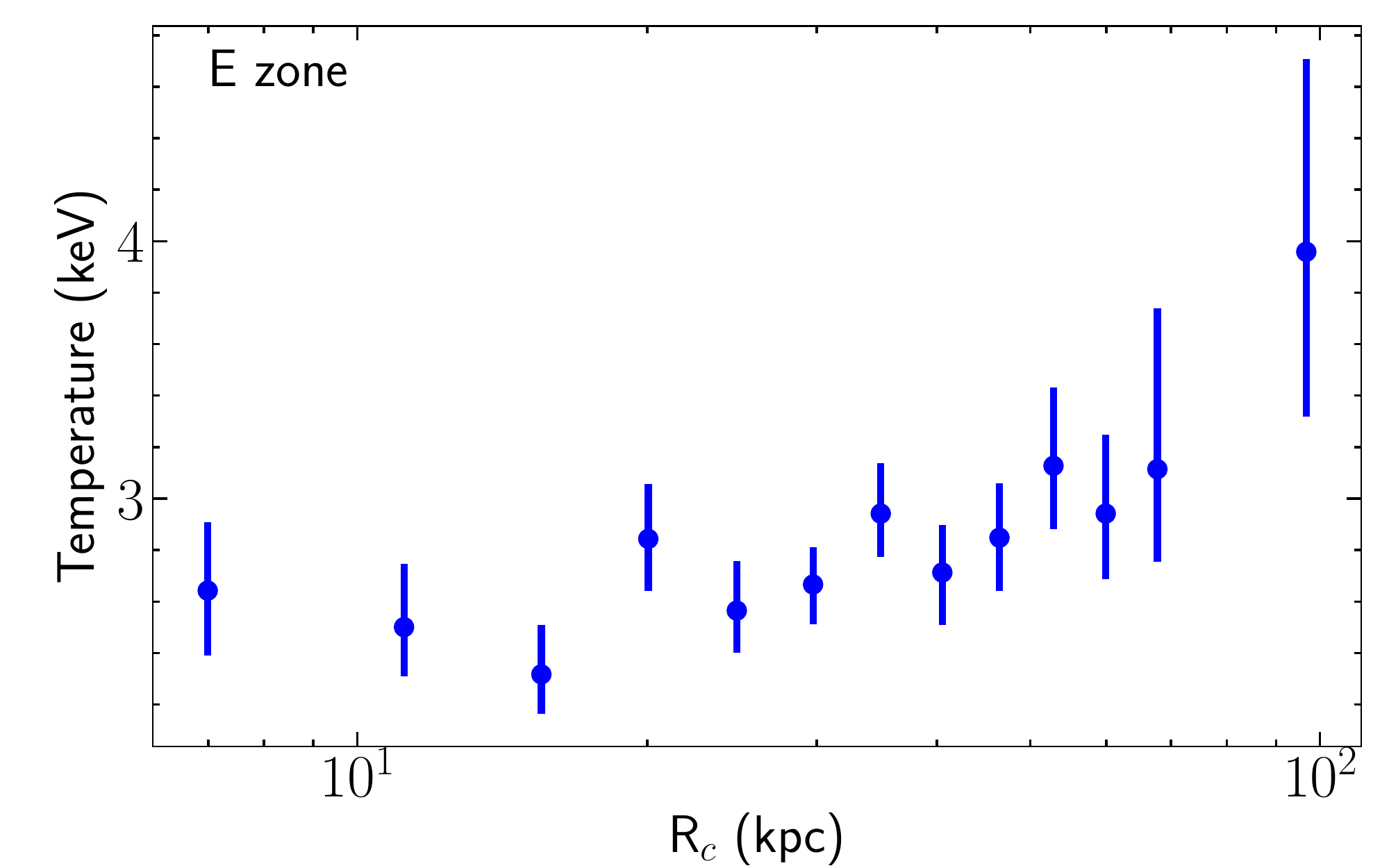}   \\
\includegraphics[width=0.47\textwidth]{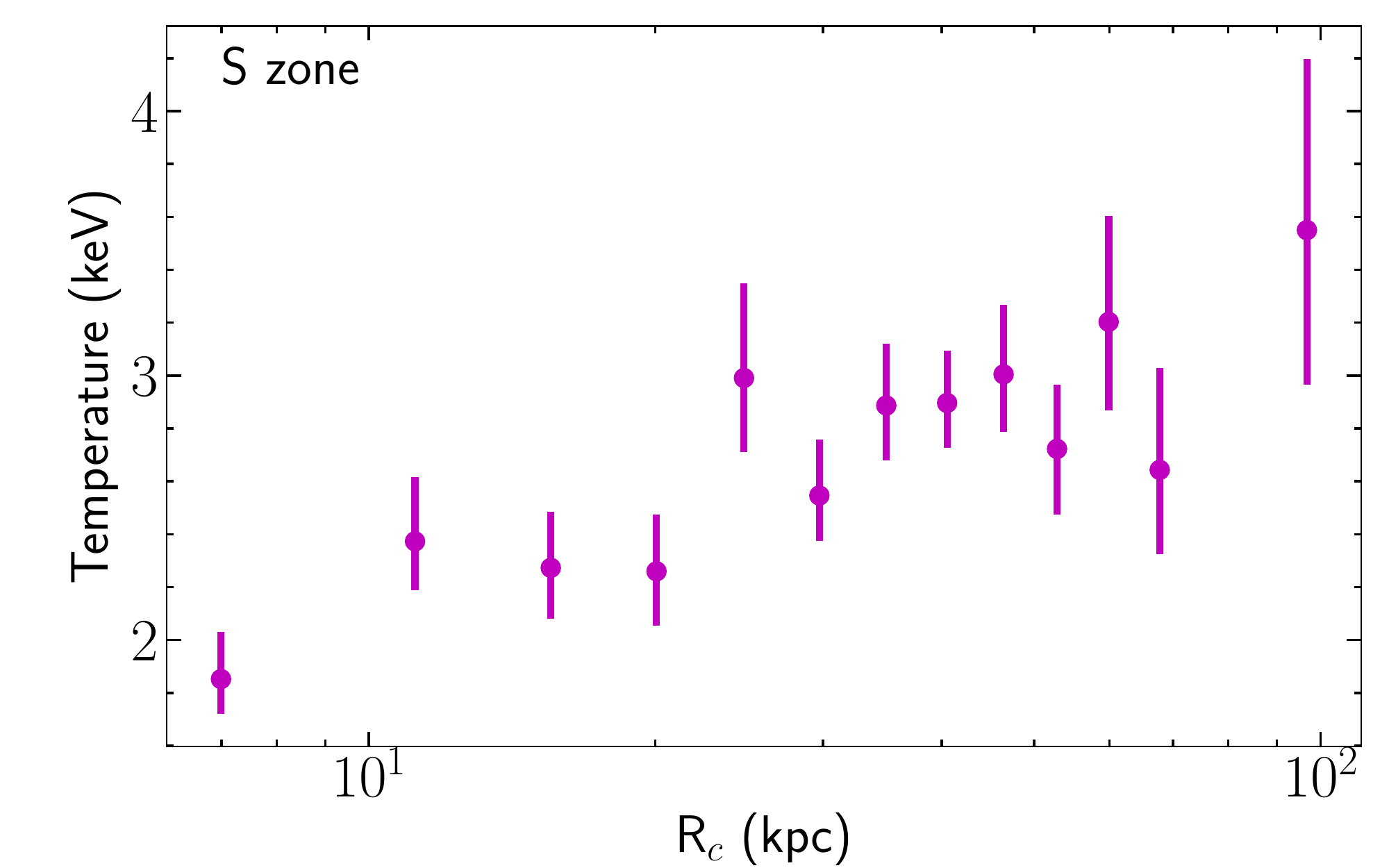}   
\includegraphics[width=0.47\textwidth]{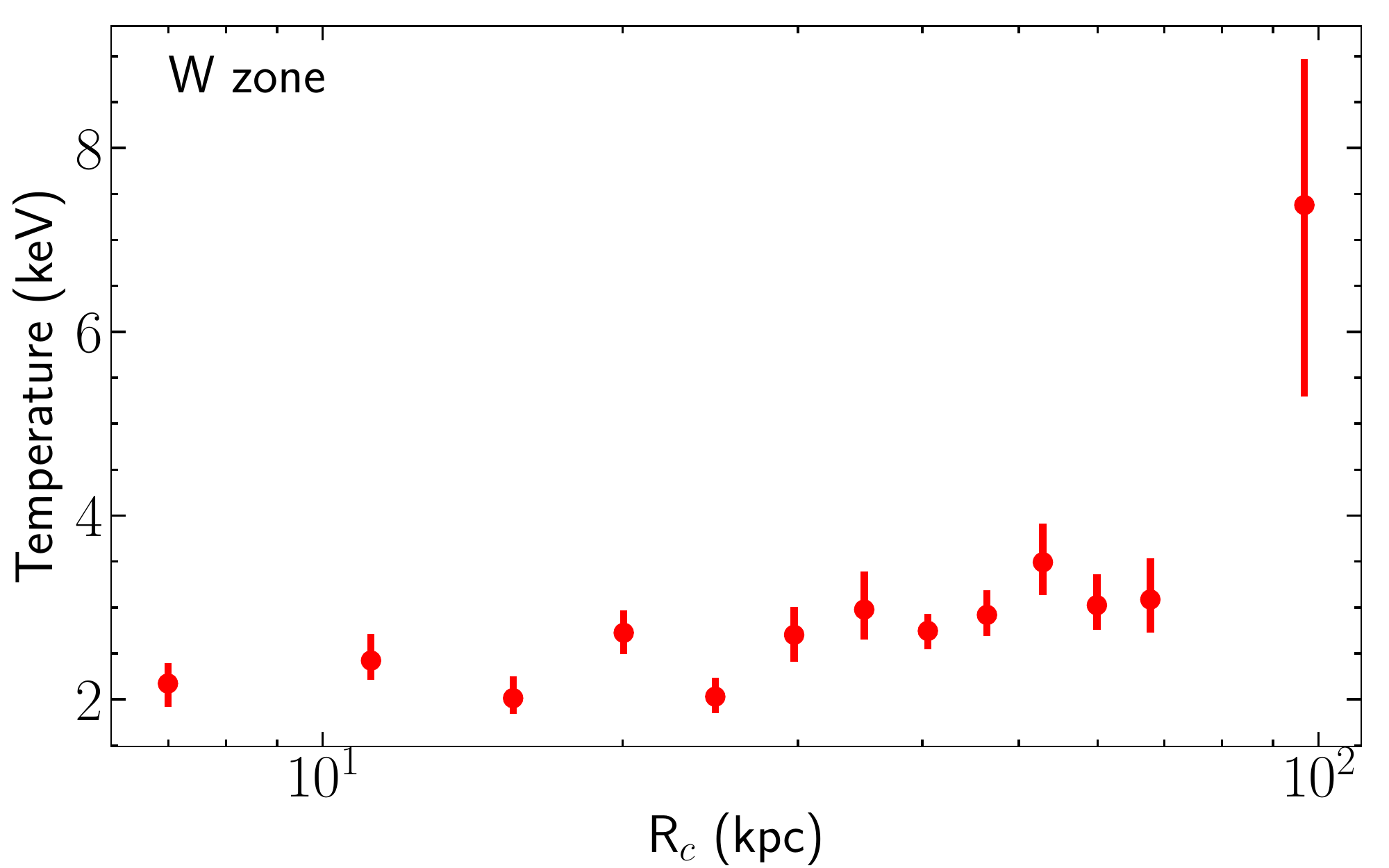}   
\caption{Temperature profiles obtained for the Virgo cluster Case 2 analysis. The colors indicate the N (green), W (red) S (magenta) and E (blue). } \label{fig_cas2_kt} 
\end{figure*}

\begin{figure*}   
\includegraphics[width=0.47\textwidth]{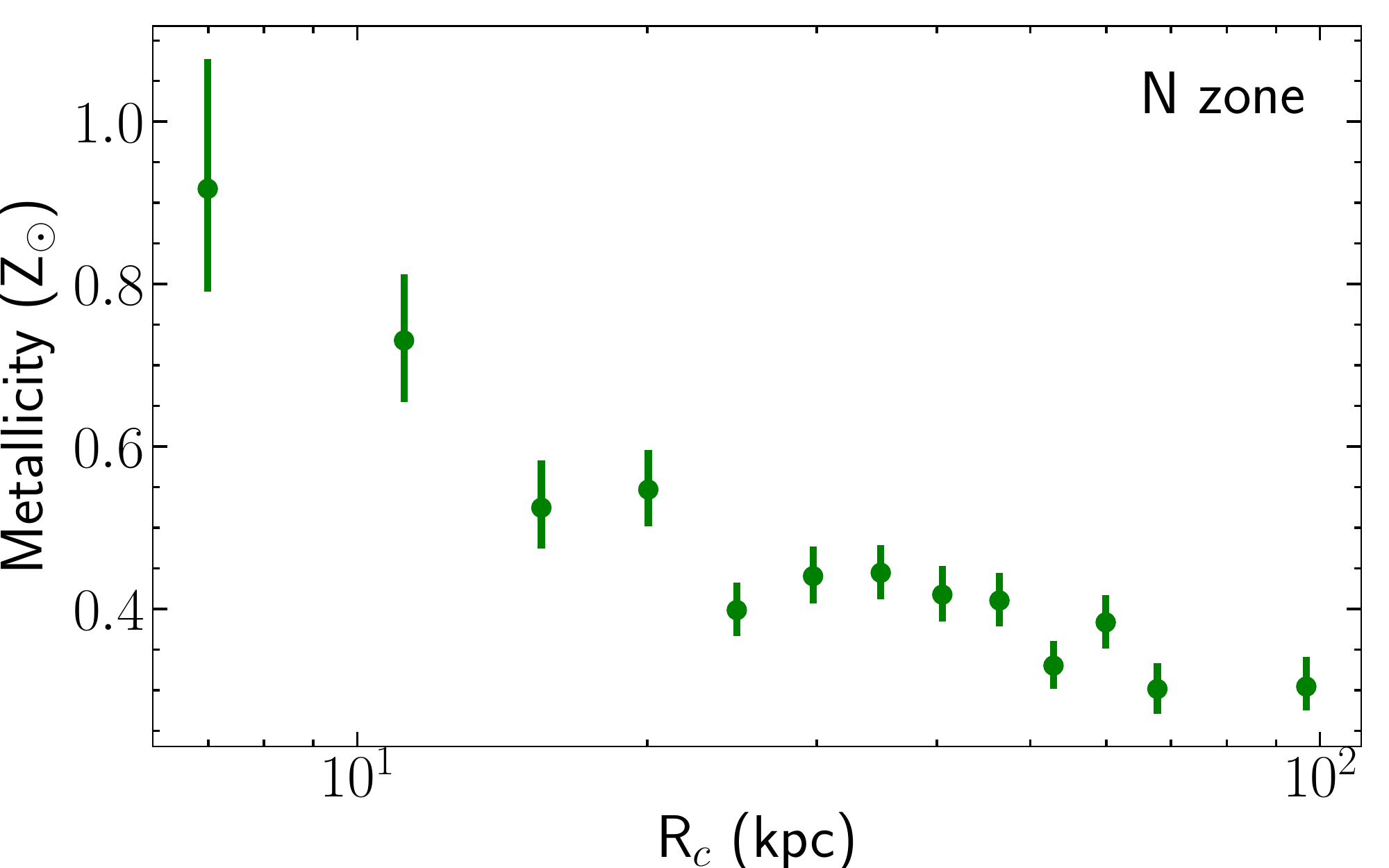}   
\includegraphics[width=0.47\textwidth]{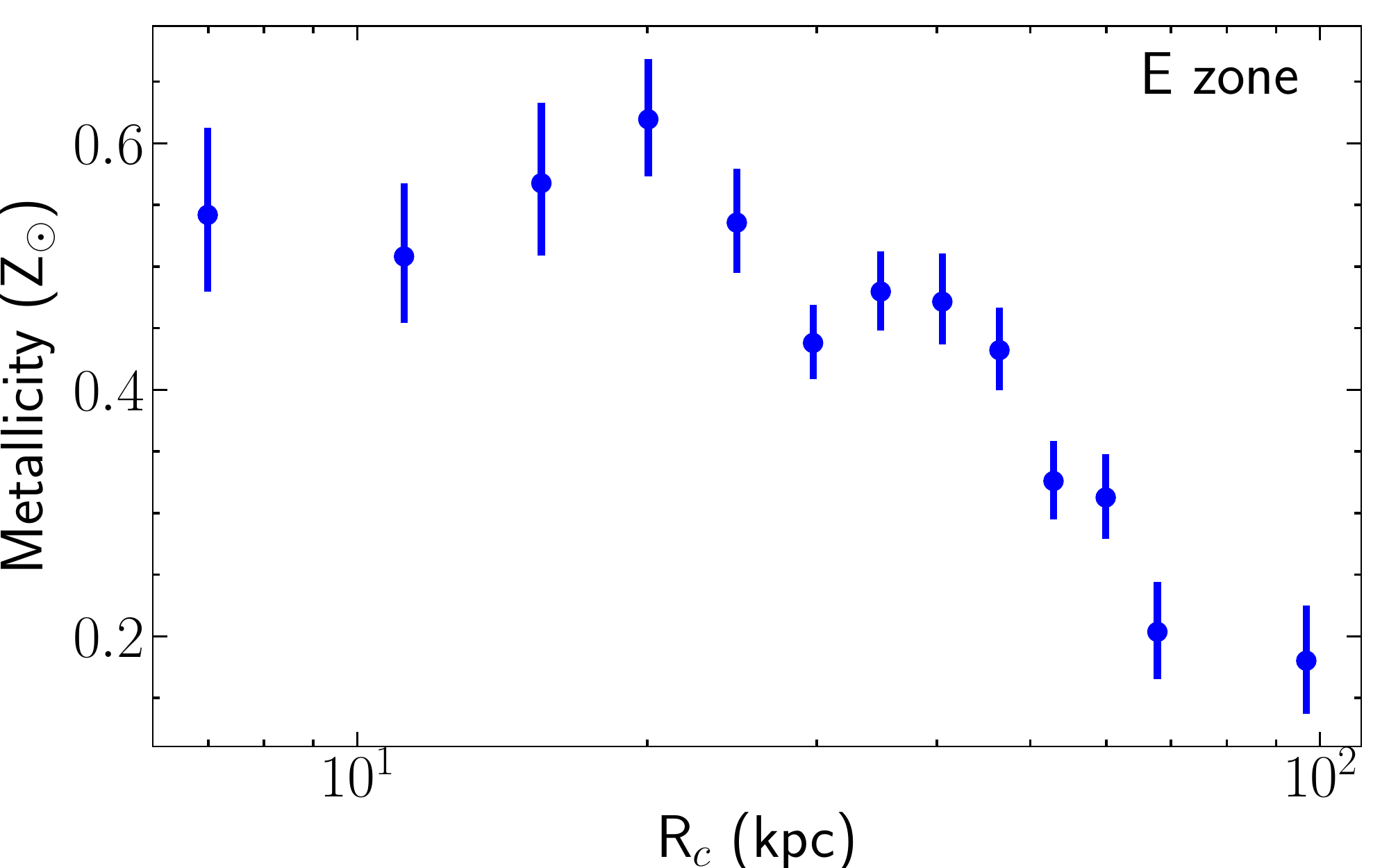}   \\
\includegraphics[width=0.47\textwidth]{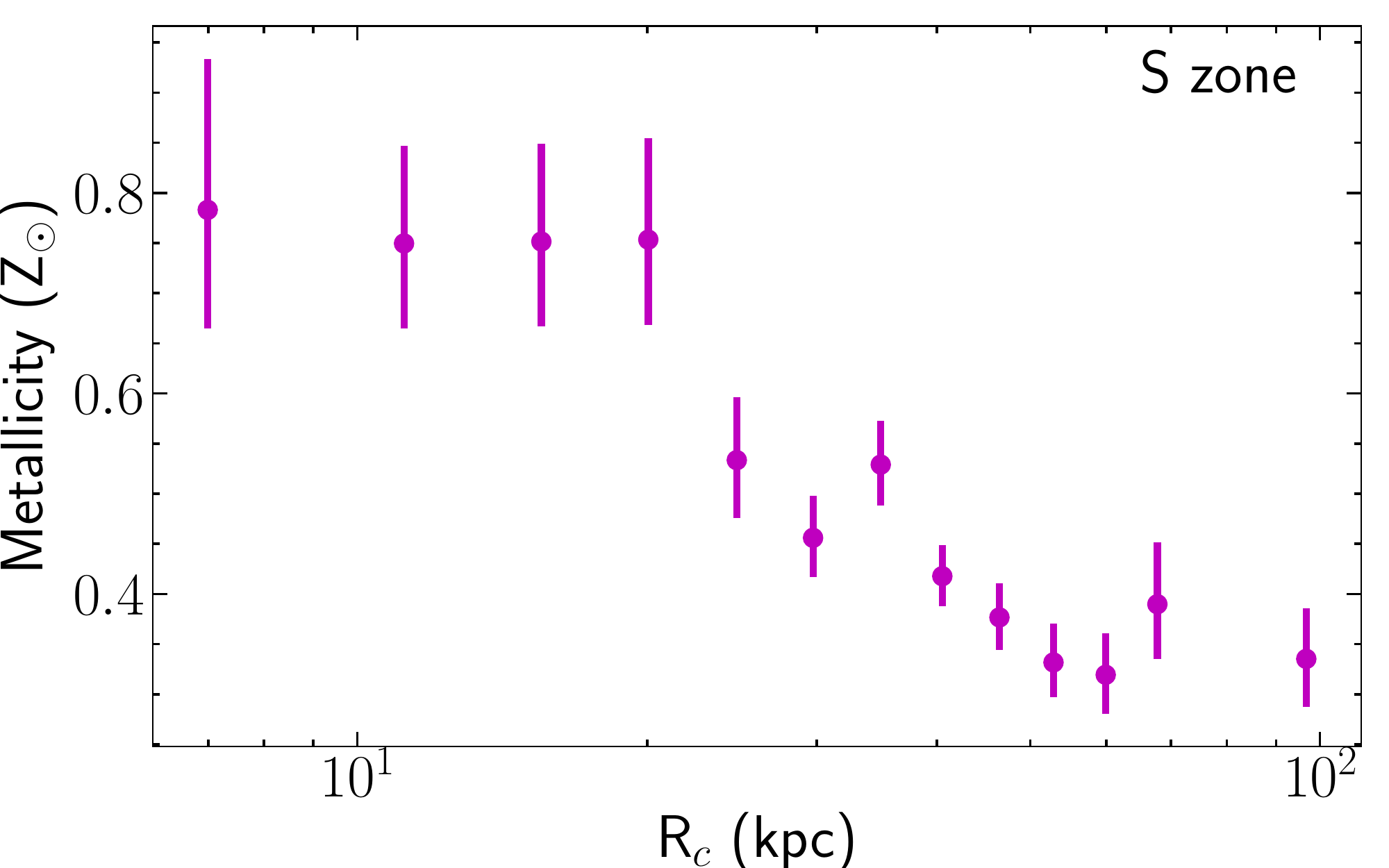}   
\includegraphics[width=0.47\textwidth]{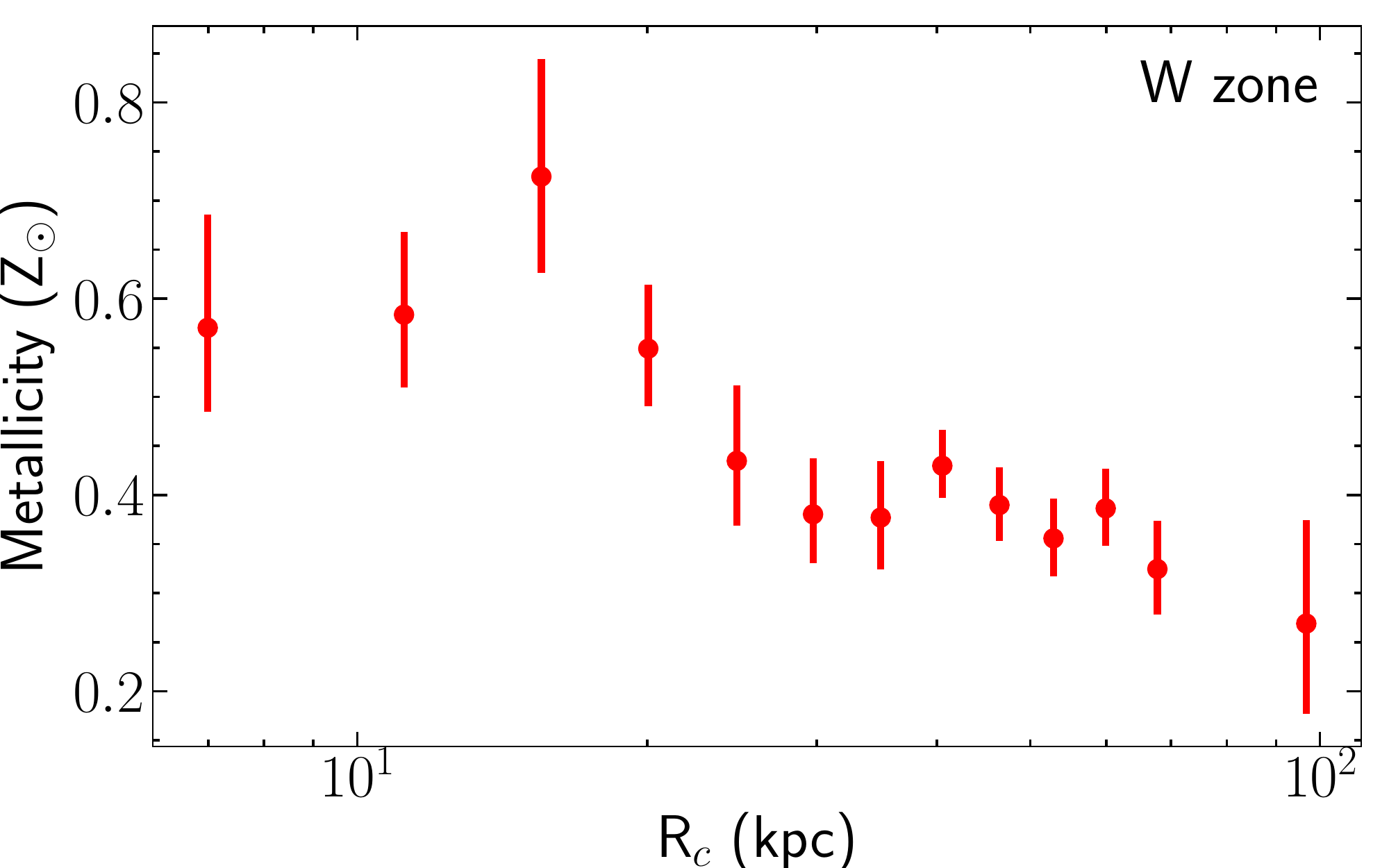}   
\caption{Metallicity profiles obtained for the Virgo cluster Case 2 analysis. The colors indicate the N (green), W (red) S (magenta) and E (blue). } \label{fig_cas2_z} 
\end{figure*}

Figure~\ref{fig_cas12_region} bottom panel shows the exact regions for \emph{Case 2}. Figure~\ref{fig_cas2_vel} shows the velocities obtained for each region (numbered from the center to the outside). A complex structure can be shown as we are closer to the cluster center. Near the center, the E region shows a redshifted structure (e.g. $1168_{-499}^{+454}$ km/s for ring 1) while the W region shows a blueshifted structure (e.g. $-1506_{-614}^{+633}$ km/s for ring 1), likely indicating the presence of an outflow near the AGN. Figures~\ref{fig_cas2_kt} and~\ref{fig_cas2_z} shows the temperature and metallicity profiles for each region. Both profiles are similar towards all zones, with temperature decreasing as we move to the cluster center while metallicity increases in the same direction. A temperature/metallicity discontinuity can be found at $\sim 10-30$ kpc towards all zones.

\subsection{Spectral maps}\label{spec_maps} 
\begin{figure*} 
\centering 
\includegraphics[width=0.485\textwidth]{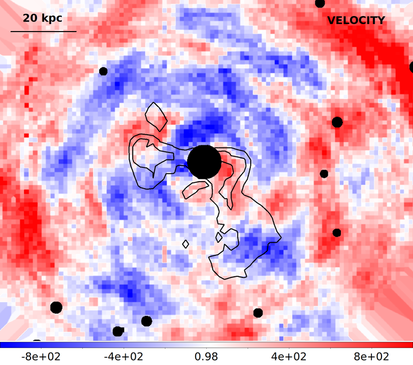} 
\includegraphics[width=0.485\textwidth]{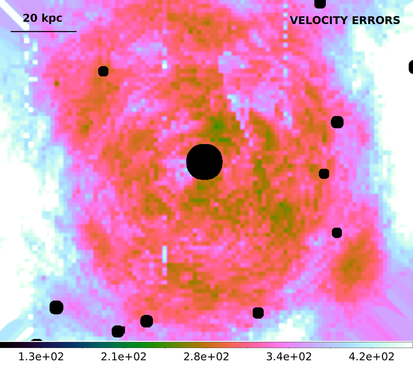} \\
\includegraphics[width=0.485\textwidth]{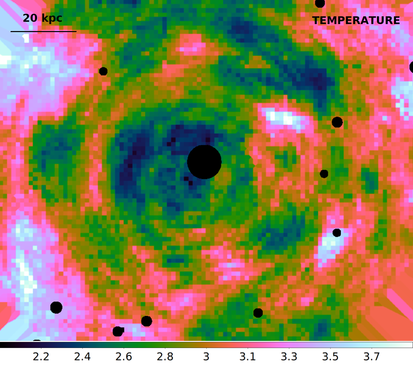} 
\includegraphics[width=0.485\textwidth]{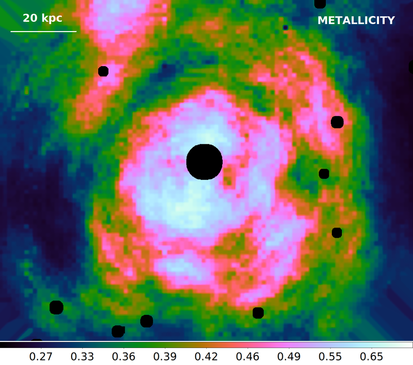}
\caption{\emph{Top left panel:} velocity map (km/s) relative to M87 ($z=0.00436$). Black circles correspond to point sources which were excluded from the  analysis, including the AGN in the cluster core. Maps were created by moving 2:1 elliptical regions (i.e. rotated to lie tangentially with respect to the nucleus) containing $\sim$500 counts in the Fe-K region. The 3-level contour map from the radio observations obtained by \citet{owe00} are included as reference for the size scale.  \emph{Top right panel:}  1$\sigma$ statistical uncertainty on the velocity map (km/s). \emph{Bottom left panel:} temperature map in units of KeV. \emph{Bottom right panel:} metallicity map relative to solar abundances from \citet{lod09}, obtained for the same elliptical regions.  } \label{fig_velocity_ellipses} 
\end{figure*}

Following \citet{san20}, we created a smooth velocity map of the cluster. We use elliptical regions with a 2:1 axis ratio, which were rotated in a way that the longest axis lay tangentially to the vector to the central core. The radii of the ellipses changed adaptively to have $\sim$ 500 counts in the Fe-K complex after continuum substraction, to reduce the uncertainties on the velocities. We moved in grid with a spacing of 0.25 arcmin. We performed a combined fit of the spectra obtained per region for all observations. For each ellipse we created weighted-average ancillary responses files (arf) created from the individual ones for each observation but used the same response matrix for all the pixels. Figure~\ref{fig_velocity_ellipses} shows the velocity maps obtained (top left and top right panels). To improve the clarity we use a blue-red color scale for the velocity map to indicate whether the gas is moving towards the observer (blue) or away (red). Also, the 3-level contour maps from the radio observations are included as reference for the size scale. The complexity of the velocity structure in the system is clear. Interestingly, we have found that the gas moves around the cluster core with different directions, following the arms identified in the soft X-ray image. The east arm displays blueshifted velocities while the southwest arm displays redshifted velocities. Also, the maps shows higher metallicity in regions perpendiculars to the soft X-ray arms near the cluster core.

\begin{figure}
   \centering
\includegraphics[width=0.46\textwidth]{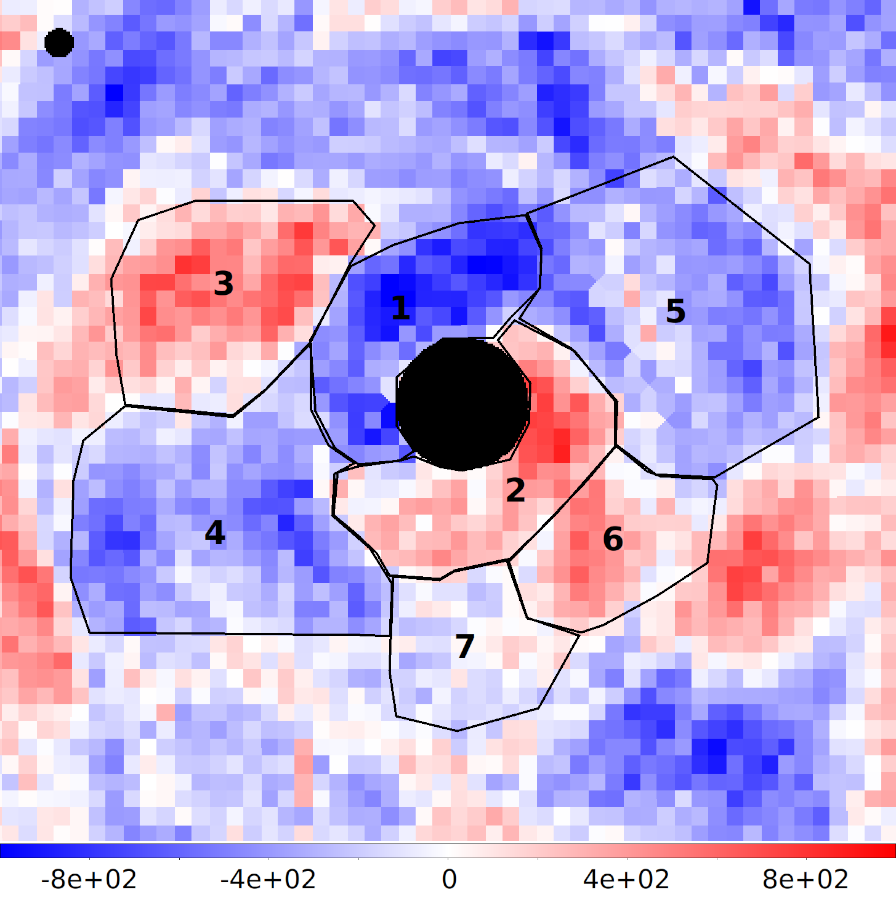}\\
\includegraphics[width=0.46\textwidth]{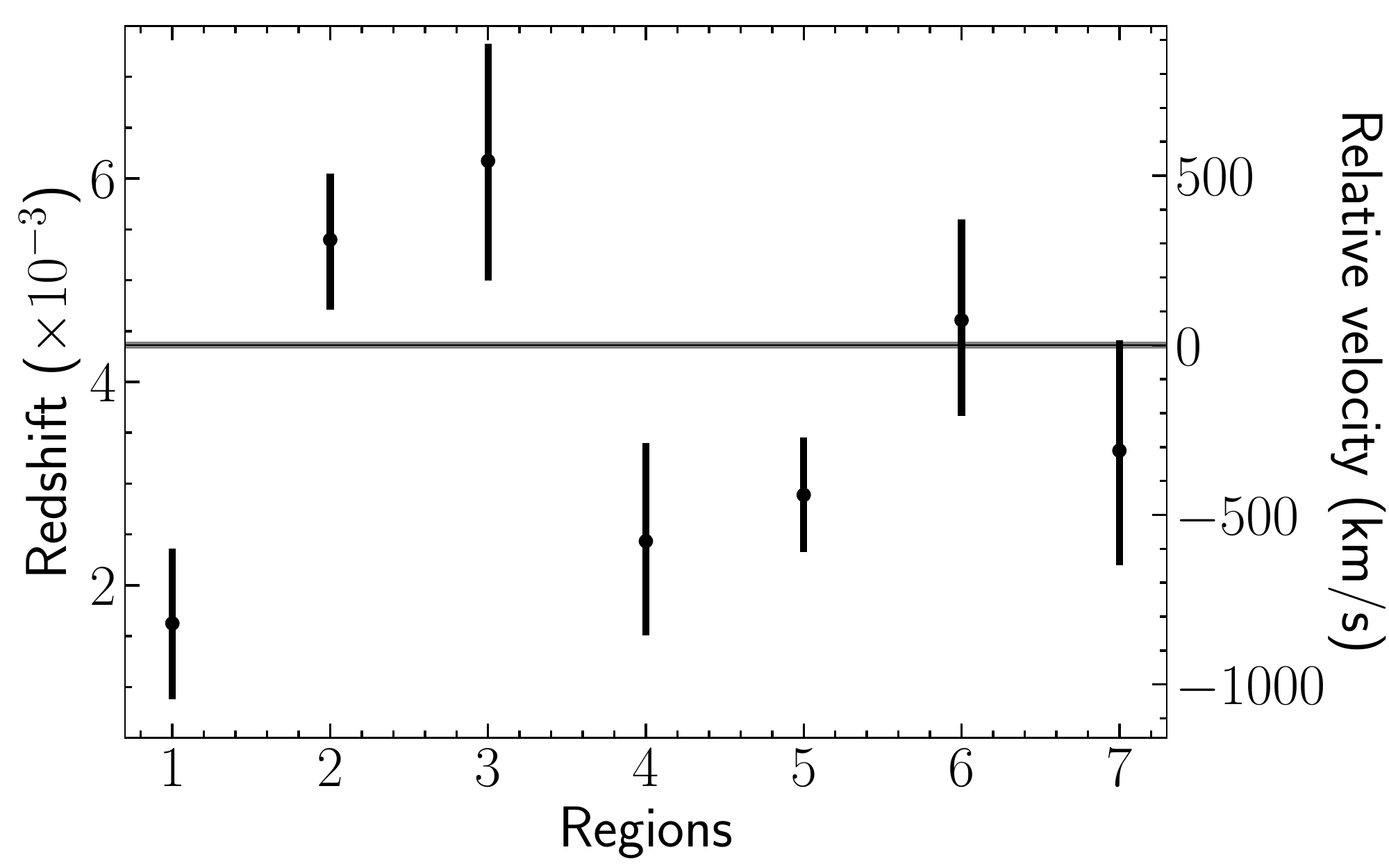}
\caption{\emph{Top panel:} Manually selected regions following the velocity spectral map. \emph{Bottom panel:} velocities obtained for each region.} \label{fig_vel_error} 
    \end{figure}

We manually analyzed selected regions close to the cluster core to analyze the redshifted-blueshifted pattern identified above. The top panel in Figure~\ref{fig_vel_error} shows the regions analyzed while the bottom panel shows the best-fit velocities. For regions 1 and 2 the velocities correspond to $-820\pm 222$ km/s and $311\pm 199$ km/s, respectively. This correspond to a departure from systemic at about the 3.8 $\sigma$ confidence level. For regions 2-5 the velocities depart from systemic by $\sim 1-2$ $\sigma$, while regions 6 and 7 seems to be consistent with the systemic velocity.  That is, we have found large difference in the gas velocities between both directions near the cluster core, which may be due to the cold front and the AGN outflows. 

\begin{figure}    
\centering
\includegraphics[width=0.47\textwidth]{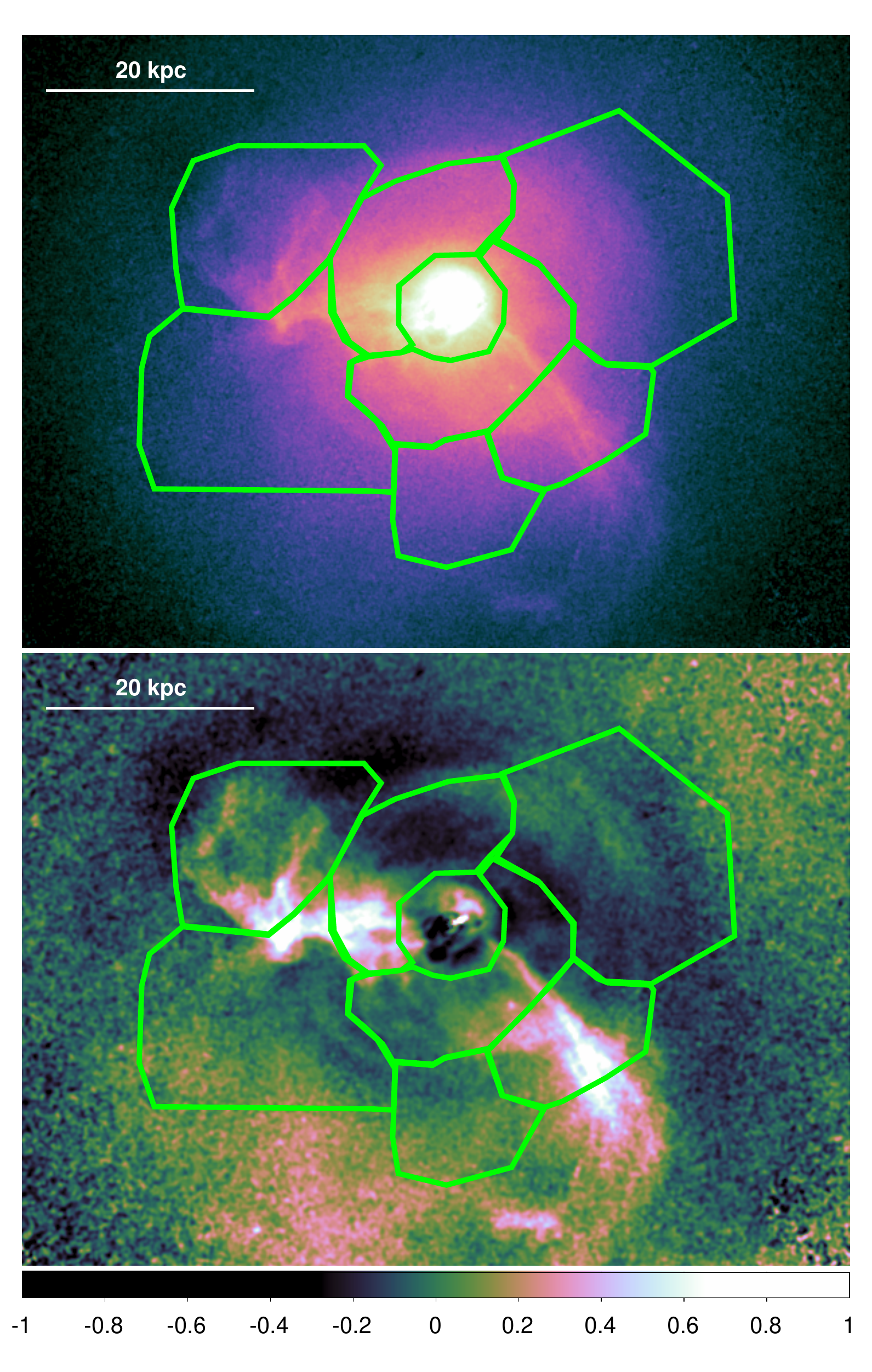}
\caption{Exposure corrected {\it Chandra} image (top panel) and fractional residual map from a symmetric model (bottom panel), showing the manually selected regions from Figure~\ref{fig_vel_error}.}\label{fig_chandra_velocity} 
\end{figure}

Figure~\ref{fig_chandra_velocity} shows these analyzed regions on the Chandra image. The top panel shows the exposure corrected and background subtracted {\it Chandra} X-ray image between 0.5 and 7.0 keV, smoothed by a Gaussian with $\sigma=1$ arcsec. Point sources were manually identified and masked out by filling their regions with random nearby pixel values before smoothing. The bottom panel shows the fractional residual map from a symmetric model. The same point-source masked image was smoothed by a Gaussian with $\sigma=2$ arcsec. The average of this map as a function of radius from the M87 nucleus was computed and then used to create a symmetric model image. Shown is the fractional difference between the smoothed image and this model.

\begin{figure} 
\centering  
\includegraphics[width=0.47\textwidth]{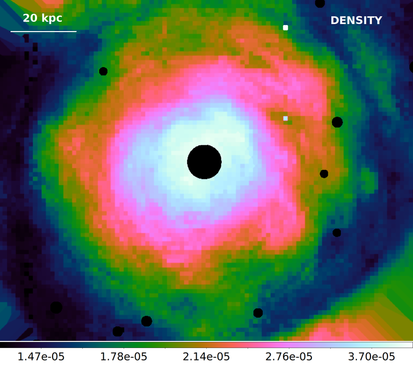} \\
\includegraphics[width=0.47\textwidth]{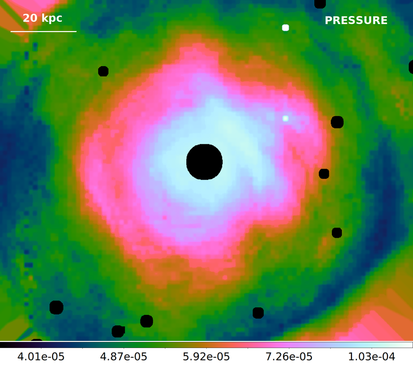} \\
\includegraphics[width=0.47\textwidth]{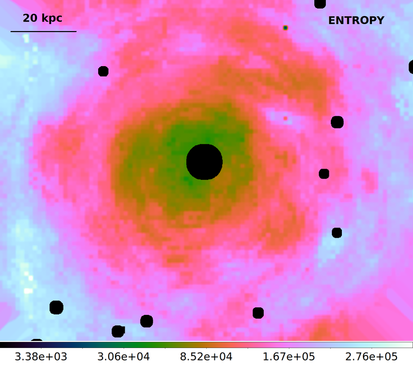}
\caption{\emph{Top panel:} density map (cm$^{-3}$). Black circles correspond to point sources excluded from the  analysis, including the AGN in the cluster core. \emph{Middle panel:} pressure map (keV cm$^{-3}$). \emph{Bottom panel:} entropy map (keV cm$^{2}$). These maps were created from the elliptical region fits described in Section~\ref{spec_maps}. } \label{fig_entropy_ellipses} 
\end{figure}

We calculated projected pseudo-density, entropy and pressure in each spatial bin using the 2D metallicity maps. We assumed a constant line-of-sight depth for all spectral regions, calculating the pseudo-density as $n\equiv \sqrt{\eta}$, where $\eta$ is the normalization of the {\tt apec} model. It is important to note that, because of the low redshift, the Virgo cluster deviates from the Hubble flow. Therefore, these pseudo-densities may be over estimated. From this we estimated pseudo-pressure as $P\equiv n\times kT$ and pseudo-entropy as $S\equiv n^{-2/3}\times kT$ \citep[see][]{hof16}. Figure~\ref{fig_entropy_ellipses} shows the density, pressure and entropy maps created from the best fit results. As expected, the region near the cluster center displays a high gas density and low entropy. The pressure plot shows a notorious ring of lower pressure in the outer region, which was identified previously by \citet{sim07}.

\subsection{X-ray/radio structures}\label{radio_region}
\begin{figure}    
\centering
\includegraphics[width=0.46\textwidth]{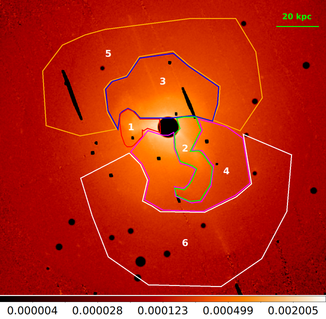}\\
\includegraphics[width=0.46\textwidth]{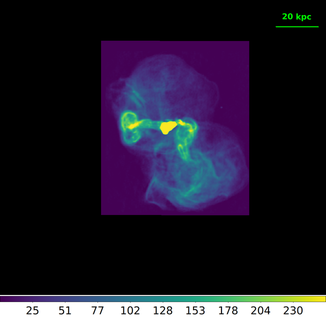}\\
\includegraphics[width=0.47\textwidth]{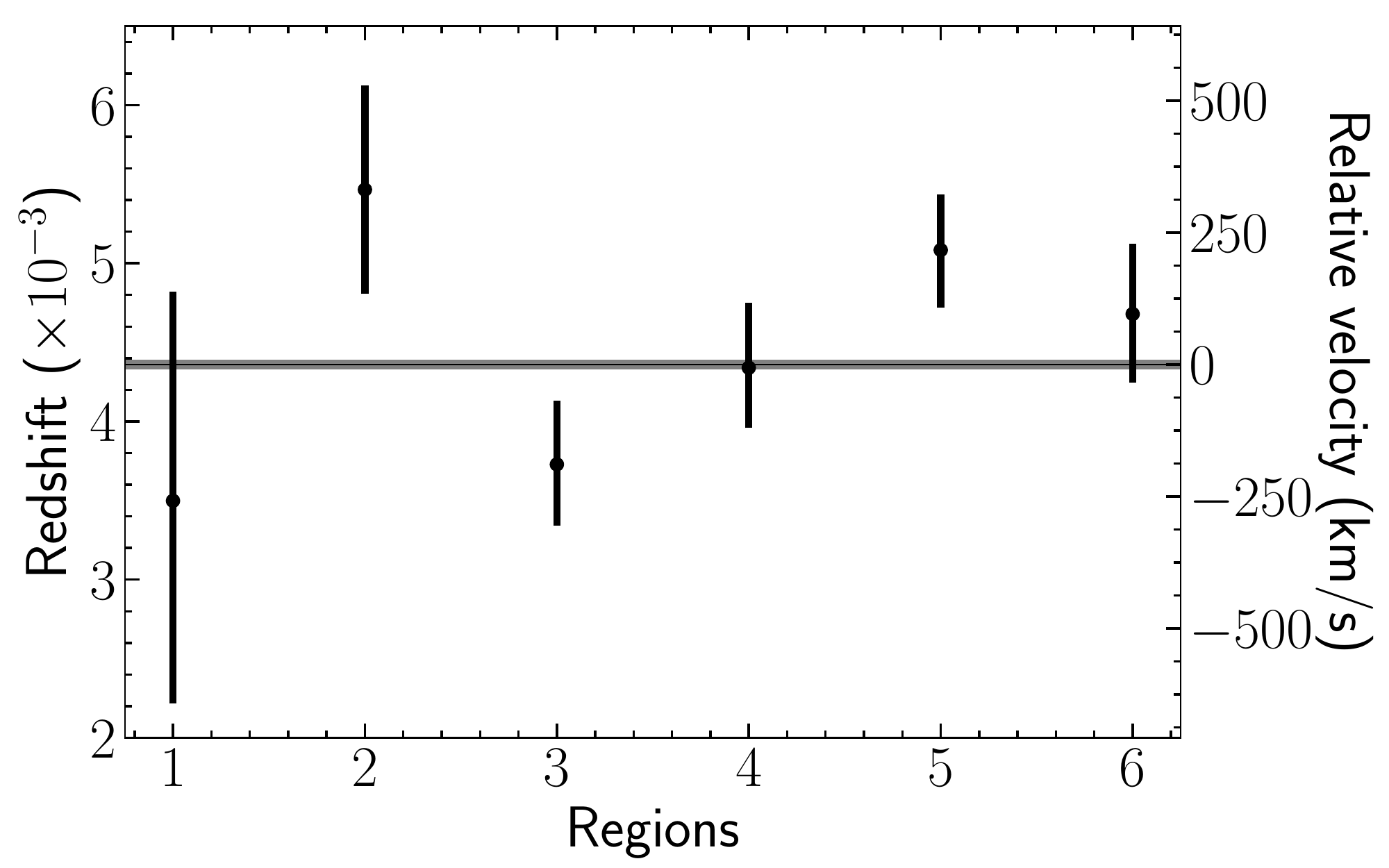}
\caption{Top panel shows the Virgo cluster extracted regions, following the radio morphological structure observed by \citet{owe00} (middle panel). Black circles correspond to point sources which were excluded from the  analysis, including the AGN in the cluster core. The bottom panel shows velocities obtained from the best-fit spectra. The M87 redshift is indicated with an horizontal line. See Section~\ref{radio_region} for more details.} \label{fig_radio1} 
\end{figure}

We have analyzed the velocity structure in non-overlaping regions following the radio morphological structure observed by \citet{owe00}. Figure~\ref{fig_radio1} shows the regions analyzed (top panel) and the radio data (middle panel). The regions were selected to cover to both eastern and western radio flows (regions 1 and 2, respectively), the radio bubbles surrounding the arms (regions 3 and 4), and two regions beyond the radio halo (regions 5 and 6). Previous {\it XMM-Newton} and {\it Chandra} observations have shown that the soft X-ray structures are  coincident with the radio flows \citep[e.g.][]{mol02,for07,sim07}. Moreover, the lack of X-ray holes at the location of the radio arms indicates that the radio plasma has a small filling factor \citep{owe00,mol02}. Regarding the size scale, note that the extracted regions shown in Figure~\ref{fig_radio1} cover a much larger region than the width of each concentric ring analyzed in Section~\ref{circle_rings}.  

Table~\ref{tab_radio} shows the best-fit results. Figure~\ref{fig_radio1} shows the velocities obtained for each region (bottom  panel). We have found velocity uncertainties down to $\Delta v \sim 100$ km/s (e.g. region 5). Interesting, the velocities obtained suggest that the gas following the eastern radio arm (Region 1) moves towards the observer, with a velocity of $-258_{-384}^{+396}$ km/s, while the gas located in the western radio arm (Region 2) is redshifted, with a velocity of $331\pm 197$ km/s. Assuming a non-relativistic plasma, \citet{owe00} found an expansion velocity for the radio halo of $160$ km/s, which corresponds to about one-quarter of the local sound speed. \citet{for07} estimate the velocity $v_{rise}$ of the buoyant bubbles identified in the {\it Chandra} soft band to be $\sim 400$ km/s. Interesting, we have found a significant change in the velocity between the gas located within the radio halo and the gas outside of it. For the western flow the velocity change is $\Delta v\sim 475$ km/s (i.e. between regions 1 and 5) while for the eastern flow the velocity change is $\Delta v\sim 236$ km/s (i.e. between regions 2 and 6).  However, it is important to note that uncertainties are large in both cases.

The velocity pattern found, i.e. redshifted towards the E outflow and blueshifted towards the SW outflow, is remarkable. Moreover, we found the same velocity structure in the spectral maps and the \emph{Case 2:} analysis (see Figures~\ref{fig_velocity_ellipses} and~\ref{fig_cas2_vel}). At small scales (< 1 kpc) it has been shown that the M87 SW jet displays both blueshifted and redshifted features along the jet \citep[e.g.][]{spa93,bir95,bri17,sni19}. Along the E direction \citet{and18} found a filament located at $\sim 2$ kpc of M87 with a blueshifted radial velocity of $-92_{-22}^{+34}$ km/s while \citet{sim18} found CO emission at $\sim 3$ kpc to the SE of M87 with a blueshifted radial velocity of $-129\pm 3$ km/s. Likewise, large scale simulations reveal the existence of outflows and backflows (i.e. hot gas flowing back to the central plane) along the jet axis up to 20 kpc of distance \citep[e.g.][]{mat08,guo11,cie17,cie18}. 

\begin{figure}   
\centering 

\includegraphics[width=0.45\textwidth]{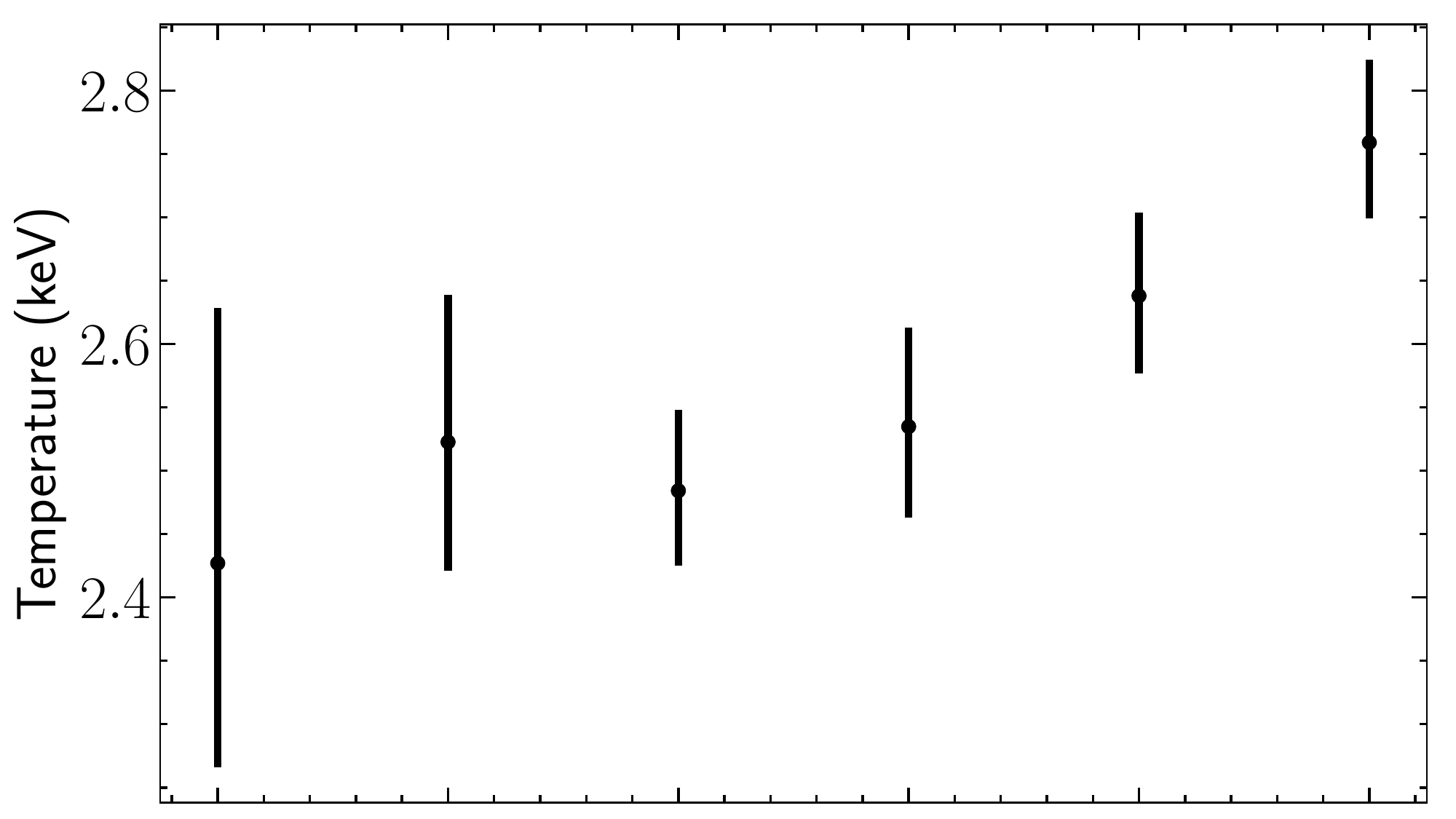}\\  
\includegraphics[width=0.45\textwidth]{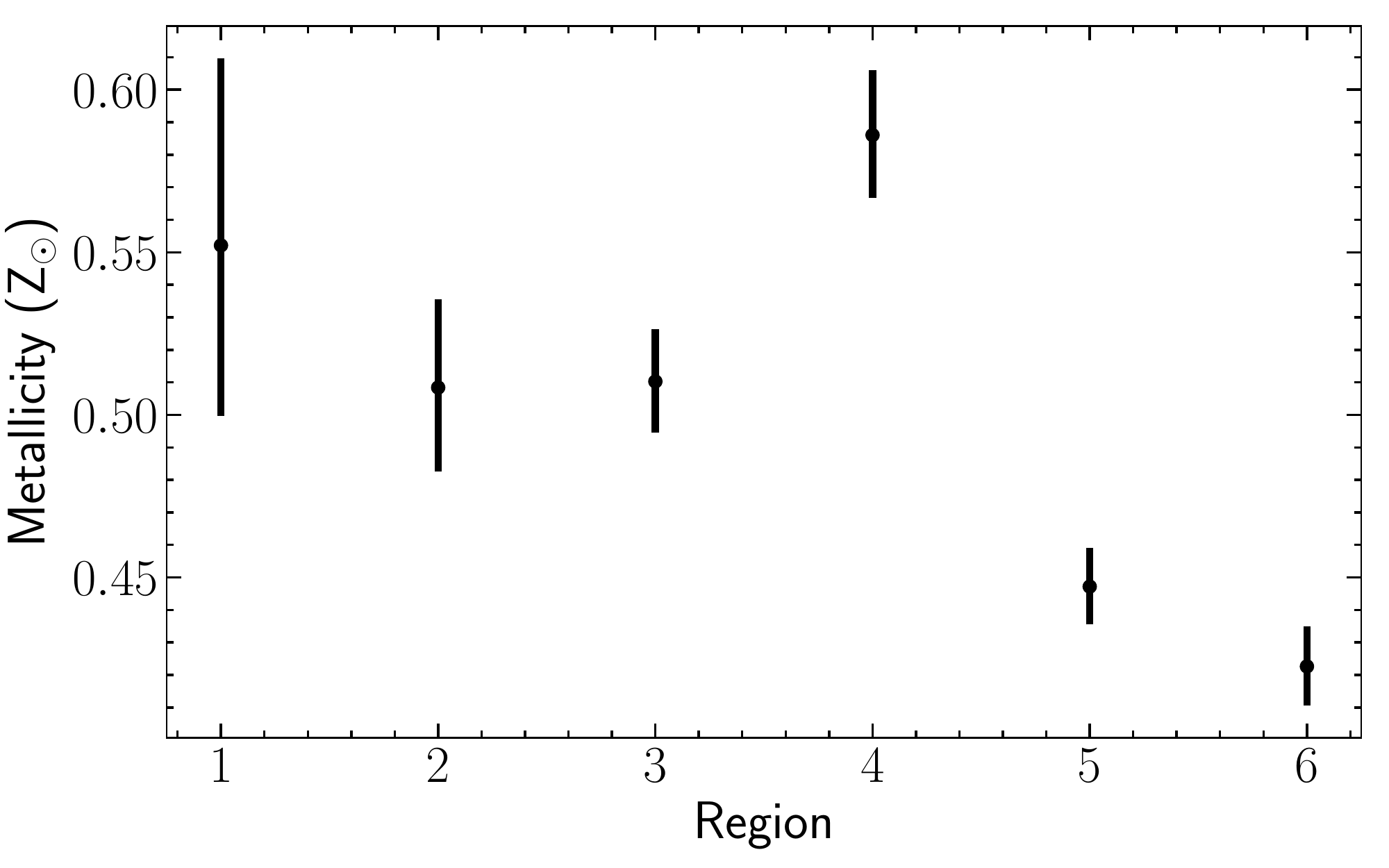} 
\caption{Temperature and metallicity profiles obtained from the best-fit spectra for the regions following the radio morphological structure observed by \citet{owe00}. See Section~\ref{radio_region} for more details.} \label{fig_radio2} 
\end{figure}

Figure~\ref{fig_radio2} shows the temperature and metallicity profiles for each region. The gas located in the eastern radio flow displays a lower temperature and slightly larger metallicity, compared to the gas located in the western radio flow, although uncertainties in bot cases are large. Interestingly, we found an enhancement in the metallicity for the region covering the south-west radio bubble in which the western radio arm is embedded (i.e. region 4). Outside the radio halo (regions 5 and 6), we have found larger temperatures with lower metallicities  for the gas with respect to the gas within the radio halo, suggesting a lack of significant metal mixing at such interface.

 \begin{table} 
\scriptsize
\caption{\label{tab_radio}Virgo cluster best-fit parameters for regions following the radio morphological structure. }
\centering
\begin{tabular}{cccccc}
\\
Region &\multicolumn{5}{c}{{\tt apec} model}  \\
\hline
 &$kT$& Z& $z$ & $norm$ & cstat/dof\\ 
  & &  &($\times 10^{-3}$) &($\times 10^{-3}$) &  \\ 
\hline
\hline
\\  
1&$2.43_{-0.16}^{+0.20}$&$0.55\pm 0.06$&$3.50_{-1.28}^{+1.32}$&$9.18\pm 0.99$&$940/982$\\
2&$2.52_{-0.10}^{+0.12}$&$0.51\pm 0.03$&$5.47\pm 0.66$&$8.53\pm 0.53$&$1222/1172$\\
3&$2.48\pm 0.06$&$0.51\pm 0.02$&$3.73\pm 0.40$&$20.81\pm 0.76$&$1391/1189$\\
4&$2.53\pm 0.08$&$0.59\pm 0.02$&$4.34_{-0.38}^{+0.41}$&$24.17\pm 1.01$&$1395/1189$\\
5&$2.64\pm 0.07$&$0.45\pm 0.01$&$5.08\pm 0.36$&$46.84\pm 1.60$&$2138/1189$\\
6&$2.76\pm 0.07$&$0.42\pm 0.01$&$4.68\pm 0.44$&$48.12_{-1.32}^{+1.36}$&$3478/1189$\\
\\ 
 \hline
\end{tabular}
\end{table}

\subsection{The NW and SE cold-fronts}\label{coldfront_region}
\begin{figure*}    
\centering
\includegraphics[width=0.45\textwidth]{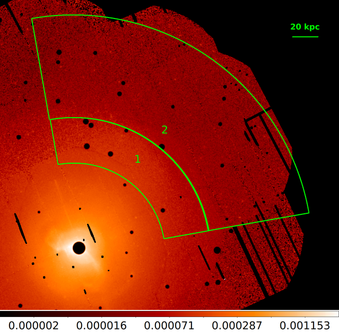}
\includegraphics[width=0.45\textwidth]{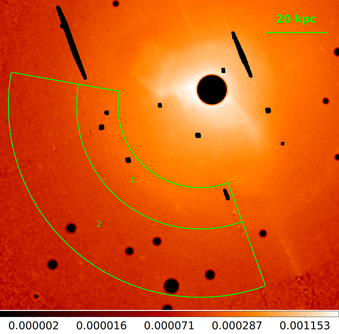}\\
\includegraphics[width=0.45\textwidth]{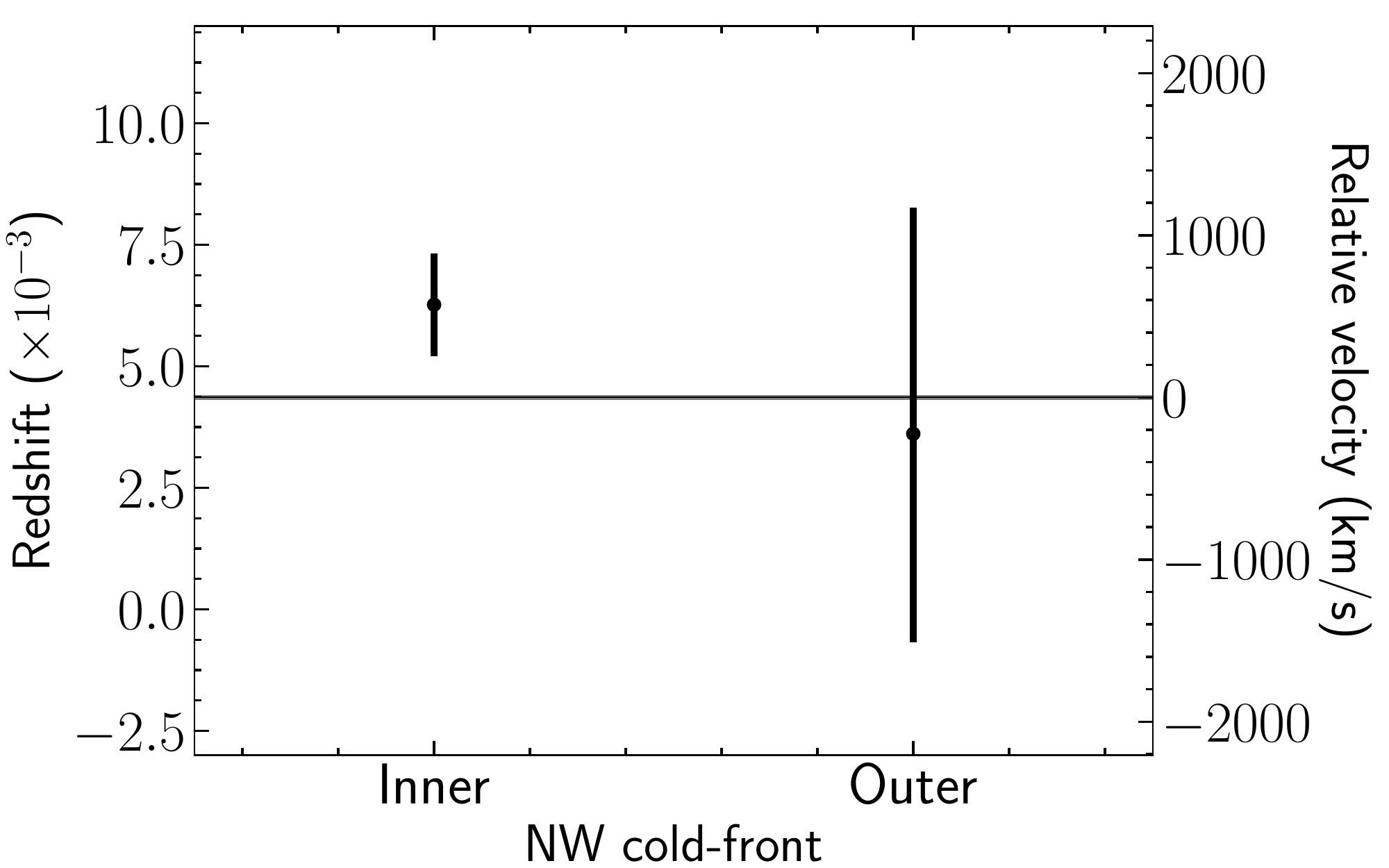}
\includegraphics[width=0.45\textwidth]{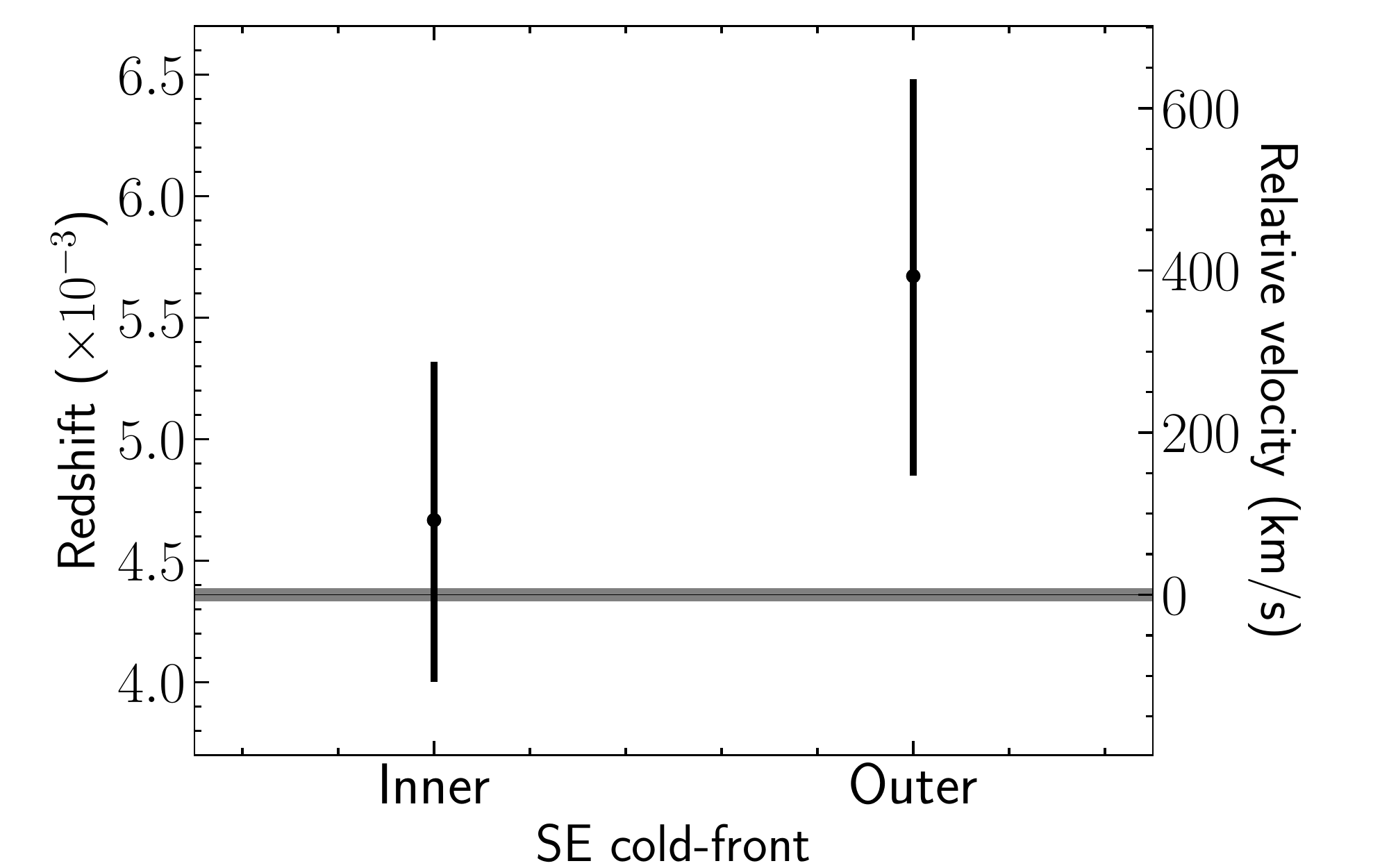}
\caption{\emph{Top panels:} Virgo cluster extracted regions, following the NW (top left panel) and SE (top right panel) cold-fronts. Black circles correspond to point sources which were excluded from the  analysis, including the AGN in the cluster core. \emph{Bottom panels:} velocities obtained from the best-fit spectra. The M87 redshift is indicated with an horizontal line. } \label{fig_coldfronts} 
\end{figure*}

We have analyzed the cold fronts identified by \citet{sim10}, one at a radius of $\sim 90$~kpc towards the NW and a second one at a radius of $\sim 33$~kpc towards the SE. We have created semi-annular regions to measure the velocity structure in both parts. Figure~\ref{fig_coldfronts} shows the NW region (top left panel) and the SE region (top right panel) analyzed, as well as the velocities, while Table~\ref{tab_coldfronts} shows the best-fit results. 

For the NW cold-front the gas in the more external region seems to be close to the rest frame, however the uncertainties are too large to provide a good constraint in the velocities while for the inner region we found a velocity of $571\pm 318$ km/s. We found that the metallicity decreases by a factor of $\sim 1.5$, a feature also identified by \citet{wer16}. For the SE cold-front we found that the gas in both regions is redshifted with velocities of $92\pm 197$ km/s (region 1) and $393\pm 245$ km/s (region 2). Similar to the NW cold-front, we have found that the metallicity in the SE cold-front decreases by a factor of $1.33$. In both cases, the temperature uncertainties for the outer region are too large to compare it with the inner region. The lack of significant metal mixing in cold fronts is likely due to its properties as a transient wave phenomena \citep{roe11}. It is important to note that the SE cold-front, being closer to the cluster center, is more influenced by the AGN.

 \begin{table}
\scriptsize
\caption{\label{tab_coldfronts}Virgo cluster best-fit parameters for regions following the NW and SE cold-fronts.  }
\centering
\begin{tabular}{cccccc}
\\
Region &\multicolumn{5}{c}{{\tt apec} model}  \\
\hline
 &$kT$& Z& $z$ & $norm$ & cstat/dof\\ 
  & &  &($\times 10^{-3}$) &($\times 10^{-3}$) &  \\ 
\hline
\hline
\\  
\multicolumn{6}{c}{NW cold-front}\\
\hline
1&$2.95\pm 0.18$&$0.27\pm 0.02$&$6.27\pm 1.06$&$24.83\pm 1.76$&$1277/1189$\\
2&$2.95\pm 0.41$&$0.18\pm 0.04$&$5.40_{-4.10}^{+4.46}$&$25.83_{-3.56}^{+5.29}$&$1232/1189$\\
\\ 
\multicolumn{6}{c}{SE cold-front}\\
\hline
1&$2.87_{-0.12}^{+0.13}$&$0.43_{-0.02}^{+0.02}$&$4.67_{-0.66}^{+0.65}$&$13.73_{-0.72}^{+0.72}$&$1246/1187$\\
2&$2.96_{-0.14}^{+0.16}$&$0.33_{-0.02}^{+0.02}$&$5.67_{-0.82}^{+0.81}$&$19.45_{-1.15}^{+1.15}$&$1269/1189$\\
\\ 
 \hline
\end{tabular}
\end{table}

\subsection{Chandra analysis with mass profile}\label{sec_chandra}
\begin{figure}    
\centering
\includegraphics[width=0.45\textwidth]{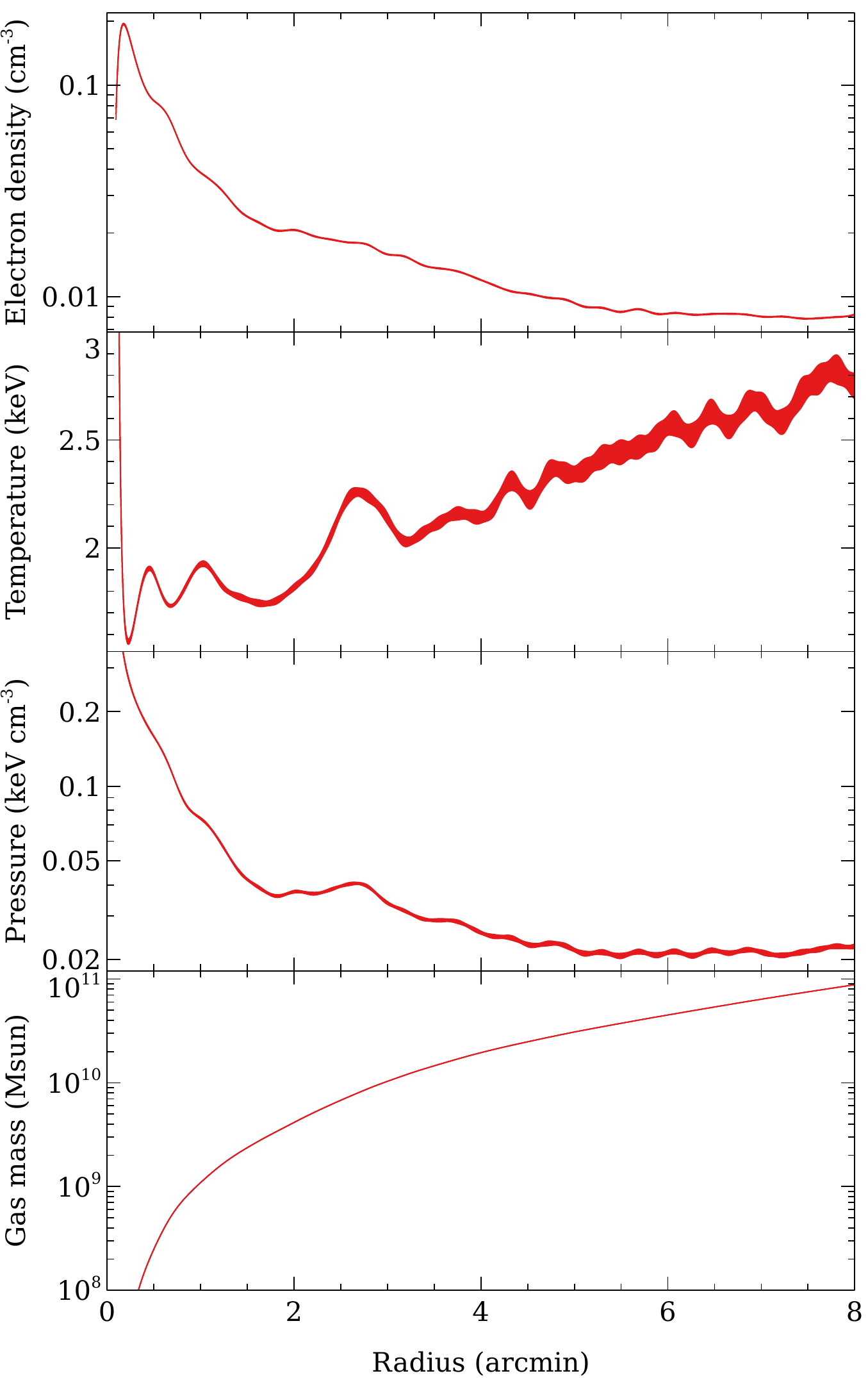}
\caption{Thermodynamic profiles for the Virgo cluster computed from the {\it Chandra} image. Uncertainties correspond to $1\sigma$ level.} \label{fig_chandra2} 
\end{figure}

We extracted profiles around the M87 nucleus from the images and exposure maps out to a radius of 8.6 arcmin, applying the point source mask. {\tt mbproj2} \citep{san18} was used to fit these profiles to compute deprojected thermodynamic profiles, with no assumption of hydrostatic equilibrium. Here the density, metallicity and temperature were modelled using spline interpolation of the values parametrized at 45 radii equally-spaced out to the maximum radius. The metallicity was also parametrized similarly using 6 radial points. The cluster model was fitted to the profiles and we used Markov Chain Monte Carlo to compute a set of model parameters. For each of these model parameters we computed the thermodynamic profiles (see Figure~\ref{fig_chandra2}). For the metallicity the inner value was frozen at the solar value due to fitting problems, possibly due to multiphase material. These profiles were analysed to compute median profiles with $1\sigma$ uncertainties. We note that there is a high temperature region close to the nucleus, despite the jet and nucleus being masked out for this analysis. This could be due to an assumption in the deprojection being incorrect, for example spherical asymmetry or multiphase plasma, or there could be a real hot component near the nucleus. The wavey structure under 2 arcmin likely reflect the edges in pressure.

\begin{figure*}    
\centering
\includegraphics[width=0.33\textwidth]{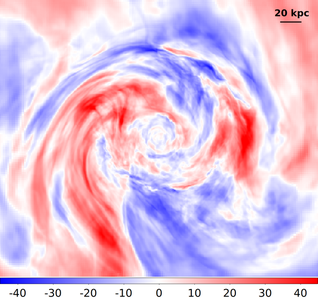}
\includegraphics[width=0.33\textwidth]{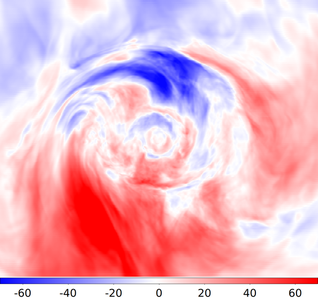}
\includegraphics[width=0.33\textwidth]{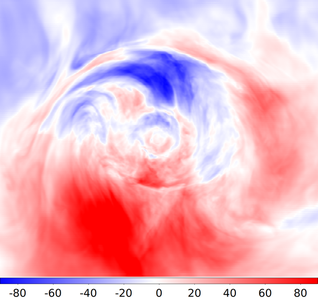}\\
\includegraphics[width=0.33\textwidth]{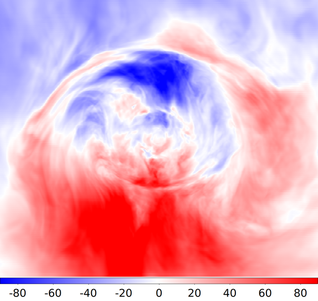}
\includegraphics[width=0.33\textwidth]{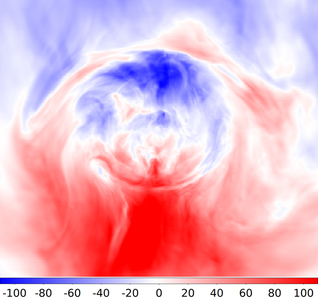}
\includegraphics[width=0.33\textwidth]{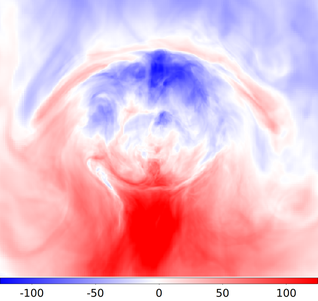}
\caption{MHD simulations of sloshing gas motions in a cluster simulated to be like Virgo. Top panels show angles of 0, 30, 45 degrees while the bottom panels show 60, 75, 90 degrees.} \label{fig_sim_mhd} 
\end{figure*}

\begin{figure} 
\centering  
\includegraphics[width=0.40\textwidth]{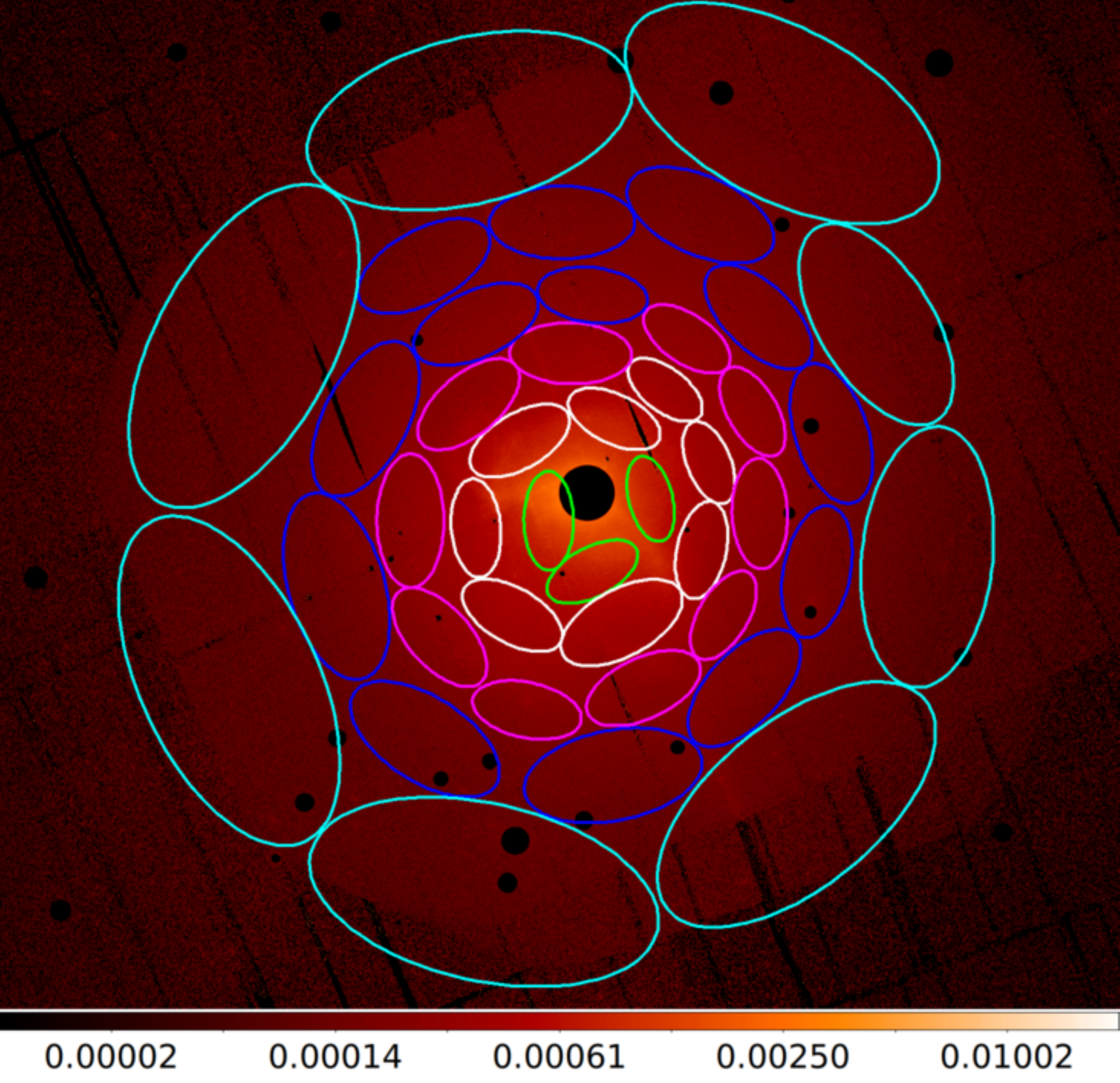}\\
\includegraphics[width=0.46\textwidth]{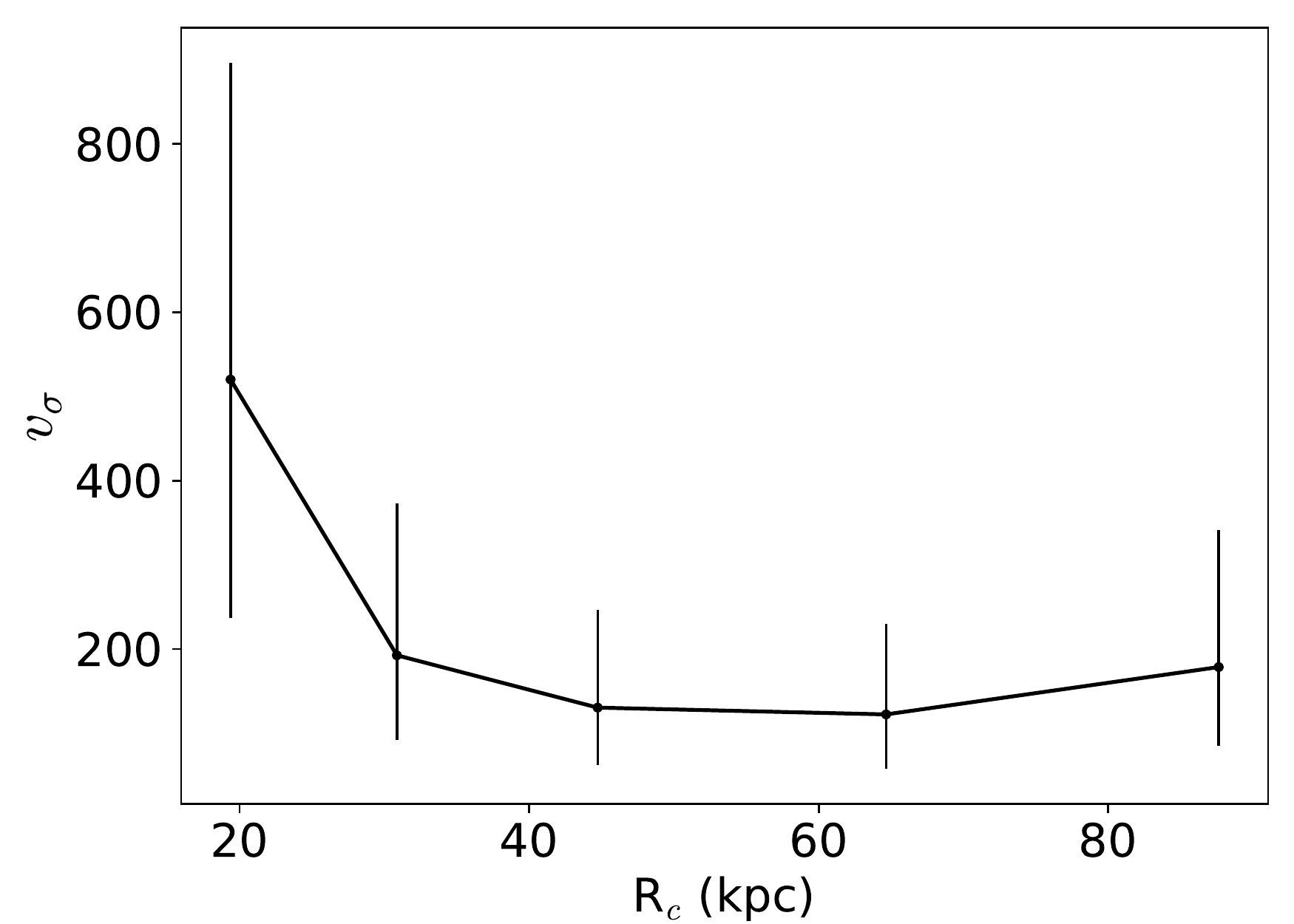} 
\includegraphics[width=0.46\textwidth]{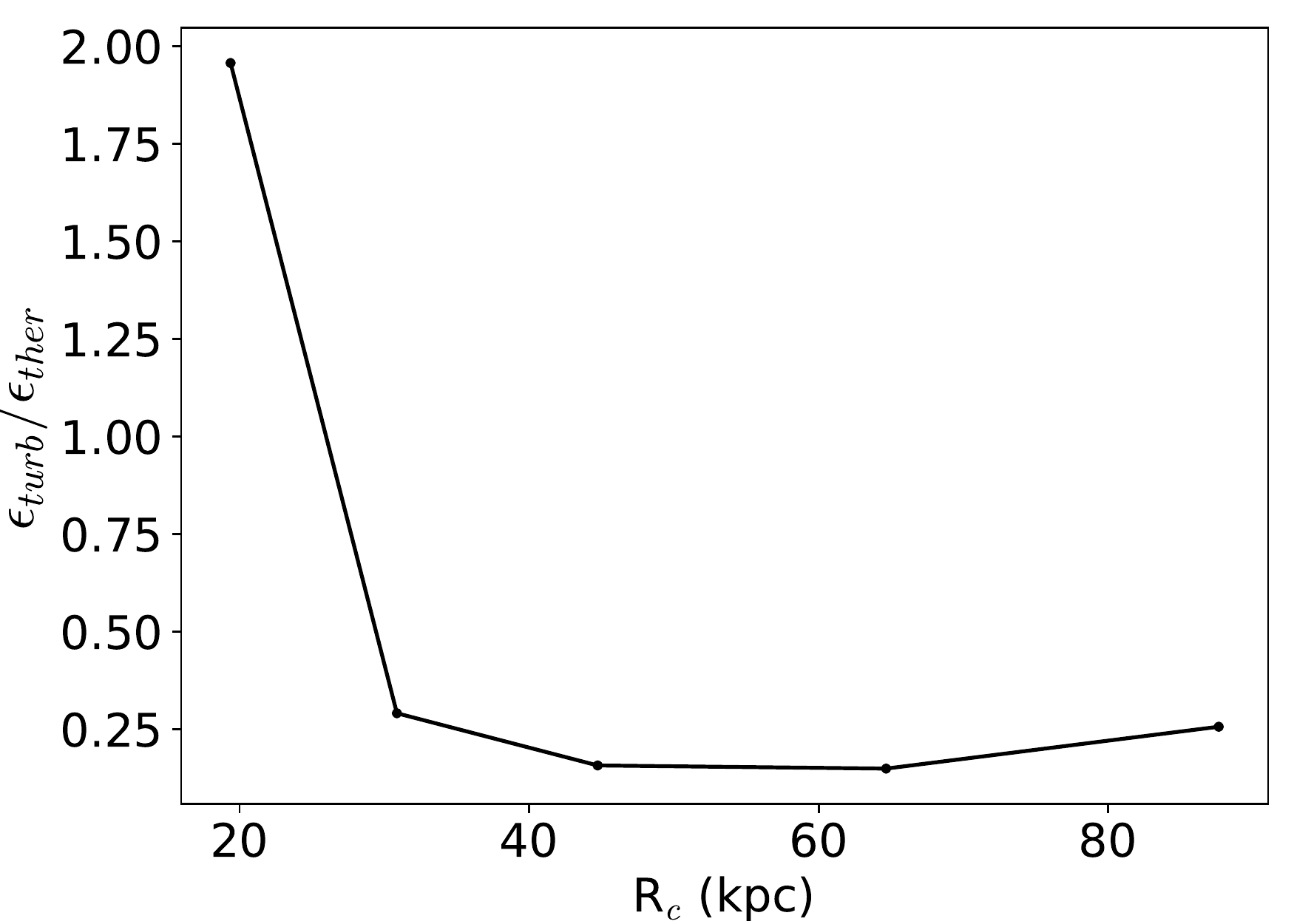}
\caption{\emph{Top panel:} non-overlapping ellipses selected to analyze the energy profile. Different colors correspond to different assigned radius. \emph{Middle panel:}  width of the velocity distribution as function of the distance from the cluster center. \emph{Bottom panel:} upper limits for the ratio of turbulent to thermal energy ($\epsilon_{turb}/\epsilon_{ther}$).}
\label{fig_ener_budget}
\end{figure}

\subsection{Systematic uncertainties}\label{sec_errors}
Apart from the energy calibration uncertainty, there are possible systematic uncertainties which may be present. With respect to the data analysis, we have found that the redshifts obtained when fitting the combined spectra (producing the results above) and when fitting simultaneously spectra from multiple observations are consistent.  Also, we have found only small variations in redshift when comparing the {\tt apec} model with the {\tt spex} CIE model (see Figure~\ref{fig_cas1_results}), suggesting that the inaccuracies in modelling is not important, when assuming a thermal plasma. 

\citet{san20} give a detailed discussion of further systematic uncertainties in their analysis of the Coma and Perseus clusters using the same technique. They have found residual calibration uncertainties at $\sim150$ km/s level, depending on the CCD spatial location. With respect to the slope for the powerlaw background component, which is fixed in the model, the choice of a powerlaw index only as a weak effect.

Finally, the maximum heliocentric velocity due to the {\it XMM-Newton} observatory motion in the solar system is $\sim 30$ km/s \citep{san20}. Such correction is relatively small, compared to the uncertainties from the spectral fits. In this sense, the large difference between blueshifted and redshifted gas identified in our analysis remains even when including such correction. 

We concluded that the above systematic effects are small compared with the measurements obtained form the best-fit spectra.

\section{Sloshing dynamics and AGN feedback}\label{sec_simul}
Our data analysis provide indications for both, AGN outflows and gas sloshing, as the origin of the complex velocity structure in the Virgo cluster. Figure~\ref{fig_sim_mhd} shows MagnetoHydroDynamic (MHD) simulations of sloshing gas motions in a cluster simulated to be like Virgo from \citet{zuh15} and \citet{wer16}. The simulation time is 2.7 Gyr, as noted in those works. Each panel shows the projected emission-weighted line-of-sight velocity in km/s, where the line of sight angle changes between the panels and is defined by a rotation from the axis perpendicular to the plane of the merger which initiated the sloshing motions (the z-axis of the simulation box) into the merger plane (along the x-axis of the simulation box). It is important to note that this simulation does not include the effects of AGN feedback. While the velocity maps do not show a clear spiral pattern, we have found an overall gradient in the velocities, with larger values as we move away from the cluster core which may indicate the effect of gas sloshing. Likewise, we have found that the velocity outside/above the NW cold front is close to the rest frame and not very well constrained, as expected from sloshing simulations where the gas motions are in the brighter, colder region. For both cold fronts we have found a drop in metallicity as we move from the inner to the outer region, a discontinuity associated to gas sloshing. 

With respect to the effects of AGN outflows, we note that the SE cold front, which is more influenced by the AGN due to its location, shows a larger velocity in the outer region compared to the inner region. While the velocity map in Figure~\ref{fig_velocity_ellipses} does not show an exact match between the velocity pattern and the radio contours, our analysis of manually selected concentric rings near the cluster center shows opposite velocities, with the gas redshifted ($1168_{-499}^{+454}$ km/s) along the E direction and blueshifted ($-1506_{-614}^{+633}$ km/s) along the W direction. Moreover, we have found changes in the velocity, metallicity and temperature in the gas located within the radio halo and the gas outside of it. Finally, the velocity pattern found in the inner region near the cluster center, with the eastern gas blueshifted and the western gas redshifted (i.e. contrary to the radio jets motion) provide hints for backflows at large scales in the gas surrounding the radio outflows.
 
In order to estimate the ratio of turbulent to thermal energy, we measured velocities for non-overlapping elliptical regions obtained in the spectral maps for different radius (See Section 3.1). The top panel in Figure~\ref{fig_ener_budget} shows the ellipse extraction regions analyzed, with different colors indicating the different radius assigned. For each radius, and assuming a gaussian distribution, we measured the velocity mean and $\sigma$-width. Middle panel in Figure~\ref{fig_ener_budget} shows the  $\sigma$-width distribution as function of the distance to the cluster center. The plot shows that the $\sigma$-width decreases as we move away from the inner radius. This is expected due to the influence of the AGN outflows near the cluster center. Using the best-fit temperature we compute the sound speed for each region as $c_{s}=\sqrt{\gamma kT/\mu m_{p}}$, where $\gamma$ is the adiabatic index, $\mu$ is the mean particle mass and $m_{p}$ is the proton mass. Then, we compute the Mach number as function of the radius. We have found that for radius $>20$ kpc the Mach number is $M \sim 0.25-0.35$, a range of values expected for gas sloshing. For the innermost radius the Mach number is $M \sim 0.89$, a value expected for AGN-driven outflow. Assuming a ratio of turbulent to thermal energy:

\begin{equation}
\frac{\epsilon_{turb}}{\epsilon_{ther}}=\frac{\gamma}{2}M^{2} 
\end{equation}

we obtain upper limits for the ratio of turbulent to thermal energy in the ICM for the different regions (see Figure~\ref{fig_ener_budget} bottom panel). For radius $>20$  kpc we found a contribution from the turbulent component $<34\%$, in good agreement with previous estimations \citep[see for example ][]{chu08,sim07,sim08,are16}.

\section{Conclusions and summary}\label{sec_con} 
Using the novel technique developed by \citet{san20} to calibrate the absolute energy scale of the {\it XMM-Newton} EPIC-pn detector we have analyzed the velocity structure in the Virgo cluster. Our results provide indications for both AGN outflows and gas sloshing, as the origin of the complex velocity structure in the Virgo cluster.  In this Section we briefly summarize our findings.

\begin{enumerate}
\item  Using this technique we have obtained accurate velocity measurements with uncertainties down to $\Delta v \sim 100$ km/s.

\item  We have created 2D projected maps for temperature, metallicity, velocity, density, pressure and entropy distribution for the Virgo cluster. 

\item  We have studied the velocity distribution by creating non-overlapping circular regions. We found that the gas located at $<30$ kpc from the cluster core displays an overall blueshifted behavior (with respect to the M87 velocity) while more external gas (i.e. $>30$ kpc) displays a redshift behavior. Such overall gradient in the velocities, with larger values as we move away from the cluster core, may indicate the effects of gas sloshing. 

\item We have analyzed the velocity distribution along N,S,W and E directions, by creating non-overlapping circular regions. We have found that the regions close to the cluster center display opposite velocities along the E and W directions, with values of $1168_{-499}^{+454}$ km/s and $-1506_{-614}^{+633}$ km/s, respectively. This may indicate the effects of the AGN outflows. 

\item We have studied the velocity structure of the gas following the morphology observed in radio observations. We have found that the hot gas located within the western radio flow is redshifted, moving with a velocity of $331\pm 197$ km/s (with respect to M87) while the hot gas located within the eastern radio flow has a velocity of $-258_{-384}^{+396}$ km/s (with respect to M87). Moreover, we have found a significant change in both, velocities and metallicities between the gas located within the radio halo and the gas outside of it. Our results may indicate the presence of backflows (i.e. hot gas flowing back to the central plane) in the region surrounding the jets.

\item We have analyzed the velocity structure in the cold fronts located towards the NW (at $\sim 90$~kpc from the cluster core) and towards the SE (at $\sim 33$~kpc from the cluster core).  Our measurements for the NW cold-front indicate that the gas located in the outer region is moving with a velocity of $571\pm 318$ km/s, while the uncertainties for the inner region are to large to provide a good constrain in the velocity (i.e. $224_{-1393}^{+1284}$ km/s). For the SE cold-front we found  that the gas moves with a velocity of $92\pm 197$ km/s and $393\pm 245$ km/s for the inner and outer regions, respectively. We note that the SE cold-front, being closer to the cluster center, is more influenced by the AGN.   

\item We analyzed the velocity distribution obtained by fitting non-overlapping regions. The width of the velocity distribution decreases as we move away from the cluster center. We have found that for radius $>20$ kpc the Mach number is $M \sim 0.25-0.35$, a range of values expected for gas sloshing. For the innermost radius the Mach number is $M \sim 0.89$, a value expected for AGN-driven outflow. For radius $>20$  kpc we found a contribution from the turbulent component of $<34\%$ to the total energetic budget.

\item We have measured the velocity of the gas in the cluster cool core using RGS data, which is in good agreement with the M87 optical velocity.
 
\item We have computed thermodynamic profiles using the {\it Chandra} images in the 0.5--7 keV energy range. We have found a high temperature region close to the nucleus, which could be due to a deprojection effect or due to the presence of a real hot component.

\end{enumerate}

Finally, the present work will be followed by a detailed study of the soft X-ray spectra as well as the Virgo cluster core.

\section{Acknowledgements} 
 The authors thank F. Hofmann for helpful discussions. This work was supported by the Deutsche Zentrum f\"ur Luft- und Raumfahrt (DLR) under the Verbundforschung programme (Kartierung der Baryongeschwindigkeit in Galaxienhaufen). This work is based on observations obtained with XMM-Newton, an ESA science mission with instruments and contributions directly funded by ESA Member States and NASA
\subsection*{Data availability}
The observations analyzed in this article are available in the {\it XMM-Newton} Science Archive (XSA\footnote{\url{http://xmm.esac.esa.int/xsa/}}).

\bibliographystyle{mnras}

%\newpage

  \appendix\label{sec_apx}

\section{XMM-Newton calibration files}\label{sec_xmm_ccf}
Table~\ref{tab_xmm_ccf}  lists the EPIC-pn calibration files used in the energy scale calibration described in Section~\ref{sec_dat_cal}.

\begin{table}  
\scriptsize
\caption{\label{tab_xmm_ccf}{\it XMM-Newton} calibration files used.}
\centering 
\begin{tabular}{lclc}   
\hline
File& Version & File & Version \\
\hline
{\tt EPN\_ADUCONV} & 0142 &{\tt EPN\_TICLOSEDODI} & 0013\\
{\tt EPN\_BACKGROUND} & 0001 &{\tt EPN\_TIMECORR} & 0015\\
{\tt EPN\_BADPIX} & 0145 &{\tt EPN\_TIMEJUMPTOL} & 0001\\
{\tt EPN\_BUCLOSEDODI} & 0005 &{\tt XMM\_ABSCOEFS} & 0004\\
{\tt EPN\_CALSOURCEDATA} & 0001 &{\tt XMM\_BORESIGHT} & 0030\\
{\tt EPN\_CTI} & 0050 &{\tt XMM\_CALINDEX} & 0218\\
{\tt EPN\_DARKFRAME} & 0001 &{\tt XMM\_MISCDATA} & 0022\\
{\tt EPN\_EFFICIENCY} & 0001 &{\tt XMM\_SPECQUAL} & 0002\\
{\tt EPN\_FILTERTRANSX} & 0019 &{\tt XRT1\_XAREAEF} & 0009\\
{\tt EPN\_FWC} & 0002 &{\tt XRT1\_XENCIREN} & 0003\\
{\tt EPN\_HKPARMINT} & 0005 &{\tt XRT1\_XPSF} & 0016\\
{\tt EPN\_LINCOORD} & 0009 &{\tt XRT2\_XAREAEF} & 0010\\
{\tt EPN\_MODEPARAM} & 0003 &{\tt XRT2\_XENCIREN} & 0003\\
{\tt EPN\_PATTERNLIB} & 0001 &{\tt XRT2\_XPSF} & 0016\\
{\tt EPN\_QUANTUMEF} & 0018 &{\tt XRT3\_XAREAEF} & 0012\\
{\tt EPN\_REDIST} & 0012 &{\tt XRT3\_XENCIREN} & 0003\\
{\tt EPN\_REJECT} & 0008 &{\tt XRT3\_XPSF} & 0018\\
\end{tabular}
\end{table}

\section{XMM-Newton spectra}\label{sec_xmm_spec}
\begin{figure*}[h]    
\includegraphics[width=1.0\textwidth]{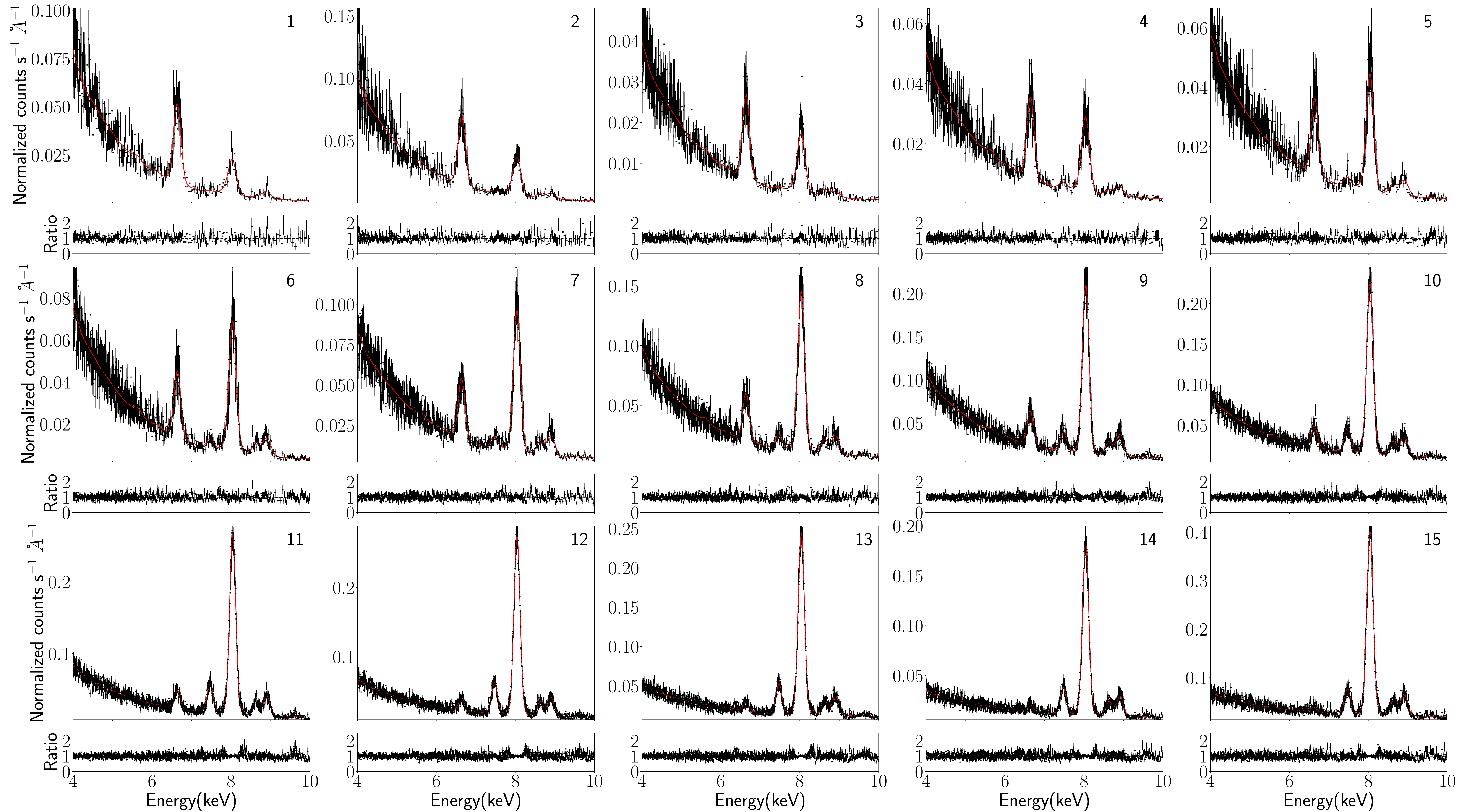}   
\caption{Best-fit spectra obtained for the Virgo cluster analysis (Case 1). The spectra have been rebinned for illustrative purposes. } \label{fig_cas1_spectra} 
\end{figure*}

As an example of the EPIC-pn spectra analyzed in this work, Figure~\ref{fig_cas1_spectra} shows best-fits spectra obtained for all regions corresponding to case 1 in Section~\ref{fig_cas1_spectra}. The spectra have been rebinned for illustrative purposes. The instrumental emission lines used as part of the background emission for the energy scale calibration are Ni K$\alpha\sim 7.5$ keV, Cu K$\alpha\sim 8$ keV,  Zn K$\alpha\sim 8.7$ keV and Cu K$\beta\sim 8.9$ keV.

\end{document}